\documentclass[a4paper,fleqn,usenatbib]{mnras}

\usepackage{mathptmx}

\usepackage[T1]{fontenc}
\usepackage{ae,aecompl}

\usepackage{mystyle}

\newcommand{\ngc}{NGC~3256}
\makeglossaries
\newacronym{aca}{ACA}{Atacama Compact Array}
\newacronym[firstplural=active galactic nuclei (AGN)]{agn}{AGN}{active galactic nucleus}
\newacronym{alma}{ALMA}{Atacama Large Millimeter/Submillimeter Array}
\newacronym{casa}{\textsc{casa}}{Common Astronomy Software Applications}
\newacronym{co}{CO}{carbon monoxide}
\newacronym{cs}{CS}{carbon monosulfide}
\newacronym{cmb}{CMB}{cosmic microwave background}
\newacronym{dec}{Dec.}{declination}
\newacronym{fov}{FoV}{field of view}
\newacronym[firstplural=full widths at half maximum (FWHMs)]{fwhm}{FWHM}{full width at half maximum}
\newacronym{gmc}{GMC}{giant molecular cloud}
\newacronym{ism}{ISM}{interstellar medium}
\newacronym{ir}{IR}{infrared}
\newacronym{kde}{KDE}{kernel density estimator}
\newacronym{lirg}{LIRG}{luminous infrared galaxy}
\newacronym{naasc}{NAASC}{North American ALMA Science Center}
\newacronym{ned}{NED}{NASA/IPAC Extragalactic Database}
\newacronym{phangs}{PHANGS-ALMA}{Physics at High Angular resolution in Nearby GalaxieS with ALMA}
\newacronym{ra}{R.A.}{right ascension}
\newacronym{rms}{RMS}{root-mean-square}
\newacronym{s/n}{S/N}{signal-to-noise}
\newacronym{ssfr}{sSFR}{specific star formation rate}
\newacronym{spw}{SPW}{spectral window}
\newacronym{tp}{TP}{total power}
\newacronym{u/lirg}{U/LIRG}{ultra/luminous infrared galaxy}
\newacronym{wmap}{\emph{WMAP}}{Wilkinson Microwave Anisotropy Probe}

\title[GMC-scale gas in NGC~3256]{Highly turbulent gas on GMC-scales in NGC~3256, the nearest luminous infrared galaxy}

\author[N. Brunetti et al.]{
Nathan Brunetti$^{1}$\thanks{E-mail: brunettn@mcmaster.ca},
Christine D. Wilson$^{1}$,
Kazimierz Sliwa$^{2}$,
Eva Schinnerer$^{2}$,
\newauthor{}
Susanne Aalto$^{3}$,
and Alison B. Peck$^{4}$
\\
$^{1}$Department of Physics and Astronomy, McMaster University, Hamilton, ON L8S 4M1, Canada\\
$^{2}$Max-Planck-Institut für Astronomie, Königstuhl 17, D-69117, Heidelberg, Germany\\
$^{3}$Department of Space, Earth and Environment, Chalmers University of Technology, Onsala Space Observatory, 43992 Onsala, Sweden\\
$^{4}$Gemini Observatory, Northern Operation Center, 67-0 N. A'Ohoku Place, Hilo, HI, USA
}

\date{Accepted XXX. Received YYY; in original form ZZZ}

\pubyear{2020}

\begin{document}
\label{firstpage}
\pagerange{\pageref{firstpage}--\pageref{lastpage}}
\maketitle

\begin{abstract}
We present the highest resolution CO (2--1) observations obtained to date ($0\farcs 25$) of NGC~3256 and use them to determine the detailed properties of the molecular interstellar medium in the central \SI{6}{\kilo\parsec} of this merger. Distributions of physical quantities are reported from pixel-by-pixel measurements at \SIlist{55;120}{\parsec} scales and compared to disc galaxies observed by PHANGS-ALMA. Mass surface densities range from \SIrange{8}{5500}{\solarmass\per\square\parsec} and velocity dispersions from \SIrange{10}{200}{\kilo\metre\per\second}. Peak brightness temperatures as large as \SI{37}{\kelvin} are measured, indicating the gas in NGC~3256 may be hotter than all regions in nearby disc galaxies measured by PHANGS-ALMA. Brightness temperatures even surpass those in the overlap region of NGC~4038/9 at the same scales. The majority of the gas appears unbound with median virial parameters of \numrange{7}{19}, although external pressure may bind some of the gas. High internal turbulent pressures of \SIrange{1e5}{1e10}{\kelvin\per\cubic\centi\metre} are found. Given the lack of significant trends in surface density, brightness temperature, and velocity dispersion with physical scale we argue the molecular gas is made up of a smooth medium down to \SI{55}{\parsec} scales, unlike the more structured medium found in the PHANGS-ALMA disc galaxies.
\end{abstract}




\section{Introduction}
Molecular gas is the immediate fuel for star formation and thus an important ingredient in galaxy evolution. In the Milky Way and nearby galaxies, \glspl{gmc} are the dominant structures within the molecular \gls{ism} \citep{Dam1987,Sol1987,Wil1990,Fuk1999,Miz2001,Eng2003,Ler2006,Ros2007} and are the sites of current and future star formation \citep{Bli1980,Gen1989,Miz1995}. However, different galactic environments exhibit molecular gas properties that vary widely while also maintaining ongoing star formation \citep[e.g.][]{Ros2005,Hey2009,Won2011,Ler2015}. A complete description of the evolution of molecular gas properties and their connections to star formation must include explanations spanning the full range of observed regimes.

Wide distributions of molecular gas mass surface densities and velocity dispersions have been observed in many regions within the Milky Way and nearby galaxies \citep[e.g.][]{Hey2001,Hey2009, Won2011,Bol2008,Col2014,Ros2005,Ler2015,Uto2015,Sun2018}. The largest and most uniform extragalactic sample was analyzed by \cite{Sun2018} including \num{11} galaxies in the \gls{phangs} survey (Leroy et al. 2020, in preparation), along with \num{4} galaxies from the literature. They found these properties vary systematically such that a narrow range of virial parameters (\numrange{\sim 1}{3}) is present. As an estimate of the balance between the kinetic energy, $K$, and gravitational energy, $U_{\mathrm{g}}$, within a molecular cloud, the virial parameter, $\alpha_{\mathrm{vir}} \equiv 2K/U_{\mathrm{g}}$, can indicate the likelihood of collapse and the ability to form stars \citep{McK1992,Fed2012,Kru2012,Pad2017}. These galaxies were also found to exhibit internal turbulent pressures ranging across \numrange{4}{5} orders of magnitude. The internal pressure, $P_{\mathrm{turb}} \propto \sigma^{2}/R$, can be estimated from observational properties of clouds and comparing the pressure to the surface density of the cloud can provide another estimate its dynamical state \citep{Ket1986,Hey2009,Fie2011,Ler2015}. These results imply the majority of molecular gas is kept close to dynamical equilibrium or collapse while the internal energy varies dramatically. Combining those molecular gas observations with measurements of the atomic and stellar surface densities, \cite{Sun2020} estimated the external pressure exerted on \glspl{gmc}. They found that maintaining dynamical equilibrium over such a large range of internal pressures requires taking into account the full external pressure exerted on \glspl{gmc} from the atomic, stellar, and clumpy molecular components of the galaxy.

While the \gls{phangs} analyses cover a range of galactic environments, they lack a direct comparison with observations of more extreme systems that bracket their sample. Do the scalings found by \cite{Sun2018} hold as gas masses, gas densities, and star formation rates continue to climb? For example, star formation efficiencies per free-fall time within \glspl{u/lirg} have been estimated to be higher than in disc galaxies. Calculated on scales of \SI{\sim 500}{\parsec} in five \glspl{u/lirg}, \cite{Wil2019} find efficiencies per free-fall time \numrange{\sim 5}{10} times higher than those found by \cite{Uto2018} for disc galaxies between \SIrange{60}{120}{\parsec}. To effect a change in the rate at which molecular gas is converted to stars, there is likely a change in the structure and dynamics of the molecular gas from which the stars form. In this work, we concentrate on the starburst regime for studying these questions.

Many \glspl{u/lirg} are gas-rich major mergers where the structure of the \gls{ism} is strongly influenced by galaxy-scale dynamics \citep{San1996,Far2001,Vei2002}. Tidal torques on the gas lead to bulk gas inflow towards the centres of mergers \citep[e.g.][]{Nog1988,Mih1996,Ion2004}. These gas motions can enhance star formation by adding additional external pressure onto \glspl{gmc}, or they could suppress star formation through the release of gravitational potential energy injecting turbulent energy into the molecular gas \citep{Kru2018}. Enhanced star formation is observed in merger systems which means that the \gls{ism} is also subjected to large amounts of clustered stellar, supernova, and cosmic-ray-heating feedback which can dramatically shape the \gls{ism} and subsequent star formation \citep{Dal2008,Kla2012,How2017,Kel2020,Boo2013,Gir2016}. The stellar activity can also pump turbulent energy into the molecular gas. The exact balance of these competing processes is not well constrained, nor is their coupling to \gls{gmc} scales in these highly active systems.

Very high angular resolution observations are required to resolve \gls{gmc} sizes of \SIrange{50}{100}{\parsec} in these distant systems. As the nearest \glsunset{lirg}\gls{lirg}, \ngc{} is a prime target to explore these effects on molecular gas in a merger for comparison with nearby disc galaxies. At a distance of \SI{44}{\mega\parsec} (\acrshort{cmb}-corrected redshift from \acrshort{ned} adopting \acrshort{wmap} five-year cosmology of $H_{0} = \SI{70.5}{\kilo\metre\per\second\per\mega\parsec}$, $\Omega = 1$, and $\Omega_{\mathrm{m}} = 0.27$), the \gls{alma} can readily resolve molecular gas on the scale of \glspl{gmc}. Its \gls{ir} luminosity is $L_{\SIrange[range-phrase=-]{8}{1000}{\micro\metre}} \SI{\sim 4e11}{\solarluminosity}$ \citep{San2003} and total star formation rate is \SI{\sim 50}{\solarmass\per\year} \citep{Sak2014}. It is a late-stage merger \citep{Sti2013} with two distinguishable nuclei separated by \SI{\sim1.1}{\kilo\parsec} \citep[][adjusted to our assumed distance of \SI{44}{\mega\parsec}]{Sak2014} that share a common envelope. The relatively face-on northern nucleus is producing a molecular outflow aimed roughly along the line of sight \citep[see diagrams in][]{Sak2014,Har2018}, which is powered by the starburst \citep{Sak2014}. The southern nucleus is nearly edge-on with an extremely collimated jet being launched in the north-south direction. Calculations by \cite{Sak2014} found that the energy budget for the southern jet likely has a contribution from a highly obscured \gls{agn}. This is consistent with \gls{ir} and X-ray observations by \cite{Ohy2015} indicating an \gls{agn} in the southern nucleus. Modeling of resolved multi-line and multi-transition molecular observations by \cite{Mic2018} also suggest the southern outflow may contain two phases produced by the interaction of the jet with the \gls{ism}.

In this paper we present the highest resolution \gls{co} observations obtained to date of \ngc{} and use them to study the detailed properties of the molecular \gls{ism}. Section~\ref{data} summarizes our observations and imaging procedure. Section~\ref{analysis} describes the steps used in analyzing \ngc{} to reproduce the methods used by \cite{Sun2018}. In Section~\ref{results}, we compare physical quantities such as surface density, velocity dispersion, peak brightness temperature, virial parameter, and turbulent pressure to those from the preliminary \gls{phangs} sample reported by \cite{Sun2018}. We discuss the implications for the structure of the \gls{ism} in \ngc{} in Section~\ref{discussion}. Section~\ref{conclusions} summarizes the conclusions and future work.

\section{Data}
\label{data}
\subsection{Observations}
Spectral line and continuum observations were carried out with the \SI{12}{\metre} main array, the \SI{7}{\metre} \gls{aca}, and the \gls{tp} antennas recovering all spatial scales down to $\sim$\ang[angle-symbol-over-decimal]{;;0.25}. Two separate configurations of the \SI{12}{\metre} array were used. Table~\ref{tab:observations} briefly summarizes the observations.

\begin{table*}
\caption{Summary of observations and calibration methods.}
\begin{threeparttable}
\label{tab:observations}
\sisetup{table-number-alignment=center,table-sign-mantissa}
\begin{tabular}{@{}l c S[table-format=1.0] S[table-format=2.0] S[table-format=4.0] c l@{}}
\hline
Array           & Observation       & {Repeats} & {Minimum}           & {Maximum}           & Calibration & \acrshort{casa} \\
                & Date              &           & {Baseline\tnote{a}} & {Baseline\tnote{a}} & Type        & Version \\
                &                   &           & {(m)}               & {(m)}               &             & \\
\hline
\SI{12}{\metre} & 1 September, 2016 & 1         & 12                  & 1713                & Pipeline    & 4.7.0 r38335; PL 38366 Cycle4-R2-B \\
\SI{12}{\metre} & 7 September, 2016 & 1         & 14                  & 2861                & Pipeline    & 4.7.0 r38335; PL 38366 Cycle4-R2-B \\
\SI{12}{\metre} & 10 March, 2016    & 1         & 13                  & 429                 & Manual      & 4.6.0 r36590 \\
\SI{7}{\metre}  & 2 November, 2015  & 1         & 9                   & 44                  & Manual      & 4.5.1 r35996 \\
\SI{7}{\metre}  & 5 November, 2015  & 2         & 8                   & 43                  & Manual      & 4.5.1 r35996 \\
\SI{7}{\metre}  & 6 November, 2015  & 2         & 7                   & 44                  & Manual      & 4.5.1 r35996 \\
\SI{7}{\metre}  & 7 November, 2015  & 2         & 8                   & 43                  & Manual      & 4.5.1 r35996 \\
\SI{7}{\metre}  & 15 December, 2015 & 2         & 7                   & 43                  & Manual      & 4.5.1 r35996 \\
\acrshort{tp}   & 5 December, 2015  & 4         & {\ldots}            & {\ldots}            & Pipeline    & 4.5.2 r36115; PL 36252 Cycle3-R4-B \\
\acrshort{tp}   & 6 December, 2015  & 2         & {\ldots}            & {\ldots}            & Pipeline    & 4.5.2 r36115; PL 36252 Cycle3-R4-B \\
\acrshort{tp}   & 15 December, 2015 & 6         & {\ldots}            & {\ldots}            & Pipeline    & 4.5.2 r36115; PL 36252 Cycle3-R4-B \\
\acrshort{tp}   & 5 March, 2016     & 1         & {\ldots}            & {\ldots}            & Pipeline    & 4.5.2 r36115; PL 36252 Cycle3-R4-B \\
\hline
\end{tabular}
\begin{tablenotes}
\item [] \emph{Notes.} All interferometric observations were multiple-pointing mosaics covering an area with a radius approximately \SI{30}{\arcsecond}, centred on \acrshort{ra}: $10^{\mathrm{h}}27^{\mathrm{m}}51^{\mathrm{s}}$ \acrshort{dec}: \ang{-43;54;15}. \Acrshortpl{spw} were centred at \SIlist{227.526;229.284;241.790;243.547}{\giga\hertz}. Each had a bandwidth of \SI{1.875}{\giga\hertz} and native spectral resolution of \SI{1.953}{\mega\hertz}.
\item[a] Projected for source position on the sky.
\end{tablenotes}
\end{threeparttable}
\end{table*}

The J=2--1 rotational transition of \gls{co} and J=\num{5}--\num{4} transition of \gls{cs} were each targeted with two overlapping \glspl{spw}. This was done to ensure the spectral line wings from the nuclear outflow and jet would be captured. Enough line-free channels were available to produce a continuum map as well.

\subsection{Calibration}
Calibration was carried out at the observatory and we produced our own cubes. \gls{casa} was used for both calibration and imaging \citep{McM2007}. Software versions used are summarized in Table~\ref{tab:observations}.

We inspected the calibration quality by plotting amplitudes and phases versus frequency, time, and $uv$ distance for all calibrator sources. We found that flagging the two antennas with the greatest separations from the array centre improved the synthesized beam pattern and \gls{rms} noise. Significant continuum emission is detected in all interferometric \glspl{spw} at the positions of the two nuclei so we performed continuum subtraction, fitting with a zeroth-order polynomial in the frequency ranges \SIrange{226.599}{227.427}{\giga\hertz}, \SIrange{229.134}{229.497}{\giga\hertz}, and \SIrange{229.759}{230.208}{\giga\hertz}.

Due to the complexity of combining observations from so many different dates, configurations, and arrays we found that performing all gridding and \gls{spw} combination with the \emph{mstransform} task, before producing cubes, gave the best results. Our criteria for quality in this step were small \gls{rms} noise variations from channel to channel and minimal discontinuities in the resultant spectra when transitioning from one \gls{spw} to the other.

Dirty \gls{co} cubes were produced from the calibrated and continuum-subtracted $uv$ data using the \emph{tclean} task in \gls{casa} version 5.4.0-68. Channels were gridded to \SI{3.906}{\mega\hertz} width (about \SI{5.131}{\kilo\metre\per\second}), starting at \SI{2004}{\kilo\metre\per\second} and ending at \SI{3599}{\kilo\metre\per\second}.

The \gls{tp} data were calibrated and imaged with the \gls{alma} single dish pipeline. The baseline subtraction step was modified to handle the fact that the \gls{co} line emission extended to one end of each \gls{spw}. Despite the modified baseline fitting, there still existed a difference of up to \num{\sim 9} per cent between the line intensities between the two \gls{co} \glspl{spw}. We tested the total flux measurement by manually stitching together the two \gls{tp} \glspl{spw} and making an integrated intensity map from this cube. We found at least \num{\sim 83} per cent of the flux was recovered in the interferometric observations using the combined \num{12} and \SI{7}{\metre} data. Given the uncertainties in the baseline subtraction process, we chose not to incorporate the \gls{tp} observations into imaging the interferometric data to avoid introducing artifacts.

\subsection{Imaging \glsentrytext{co}}
To help guide the clean algorithm in modeling emission on both scales larger than the synthesized beam and point sources, we ran multi-scale and point-source cleaning in two separate passes. The first pass was a shallow multi-scale clean using very extended hand-drawn clean masks and excluding point-source models. The second pass continued cleaning but this time using only point-source models. New masks were produced in the second pass because only compact emission remained and new masks were made to closely follow those features. Masking was accomplished primarily with the automultithresh algorithm, but we also manually edited the masks at each major cycle.

\section{Analysis}
\label{analysis}
Following \cite{Sun2018}, we convolved the dirty and cleaned \gls{co} cubes to have circular synthesized beams with \glspl{fwhm} of \SI{55}{\parsec}, \SI{80}{\parsec}, and \SI{120}{\parsec}. Pixels in the convolved cubes were resampled to a coarser grid of square pixels that were \gls{fwhm}/\num{2} on a side, thus Nyquist sampling the beams.

To derive physical properties of the molecular gas traced by the \gls{co} observations we calculated moment \num{0} (integrated intensity) and \num{2} (intensity-weighted velocity dispersion) maps. Signal masks for moment map calculations were produced following the procedure\footnote{The \textsc{Python} script for producing the signal masks were obtained from \url{https://github.com/astrojysun/Sun_Astro_Tools/blob/master/sun_astro_tools/spectralcube.py}.} used by \cite{Sun2018} (hereafter referred to as the Sun thresholding), and we refer the reader to their description of the algorithm outlined in their Section~\num{3.2}. These signal masks were then used in producing moment maps with the standard functions in the spectral-cube library \citep{Gin2019}. The Sun thresholding appeared superior to standard $n$-$\sigma$ thresholding due to its conservative treatment of the outer regions of our field of view, its ability to extract low \gls{s/n} emission without introducing significant numbers of noisy pixels, and being the only method that made the northern outflow clearly visible in the moment \num{2} map.

To convert the integrated intensity maps to mass surface density units we adopt the \gls{u/lirg} \gls{co}-to-$\mathrm{H}_{2}$ conversion factor \citep{Dow1998}, including a factor of \num{1.36} to account for helium. Following the procedure of \cite{He2020} to determine the appropriate conversion factor we use the stellar mass of \ngc{} from \cite{How2010} of \SI{1.14e11}{\solarmass} and estimate the expected \gls{ssfr} using the xGASS star-forming main sequence fit $\log{\mathrm{sSFR_{MS}}} = -0.344(\log{M_{\star}} - 9) - 9.822$ from \cite{Cat2018}. Using the $M_{\star}$ and star formation rate we calculate the ratio of the actual \gls{ssfr} to the main sequence expected \gls{ssfr} to be \num{15}. This ratio puts \ngc{} well away from the star-forming main sequence locus indicating a probability near \num{100} per cent of \ngc{} being in a starburst phase \citep{Sar2014}. The conversion factor recipe from \cite{Vio2018} is $\alpha_{\mathrm{CO}} = (1 - f_{\mathrm{SB}}) \times \alpha_{\mathrm{CO,MS}} + f_{\mathrm{SB}} \times \alpha_{\mathrm{CO,SB}}$, where $f_{\mathrm{SB}}$ is the probability of a galaxy being in a starburst phase. $\alpha_{\mathrm{CO,MS}}$ and $\alpha_{\mathrm{CO,SB}}$ are the main sequence and starburst conversion factors expected in the 2-Star Formation Mode framework of \cite{Sar2014}. Using this recipe we estimate that the conversion factor would be dominated by the starburst value so we adopt it for simplicity.

We adopt the mean ratio \gls{co} J=\num{2}--\num{1} / \num{1}--\num{0} = \num{0.79} from the xCOLD GASS sample \citep{Sai2017} which is consistent with the ratio of \num{0.8 \pm 0.22} measured in \ngc{} by \cite{Aal1995} in a single-dish beam roughly the size of our \gls{fov}. This ratio results in a conversion factor from the integrated \num{2}--\num{1} line to molecular gas mass of $\alpha_{\mathrm{CO}} = $\SI{1.38}{\solarmass\per\square\parsec}(\si{\kelvin\kilo\metre\per\second})$^{-1}$. The \num{2}--\num{1} / \num{1}--\num{0} ratio may be larger in \glspl{u/lirg} \citep{Pap2012,Sait2017,He2020} than the normal spirals studied by xCOLD GASS, which would result in the surface densities presented here overestimating the true surface densities. It is also likely there is considerable variation in the line ratio within \ngc{} (Harada et al. in preparation).

As in \cite{Sun2018}, we corrected the velocity dispersion estimates measured in the moment \num{2} maps for the finite channel widths using Equations~\numrange{15}{17} from \cite{Ler2016}. To measure the correlation between channels we calculated the Pearson correlation coefficient between each channel (``x-value'') and the next channel (``y-value''). The correction is always less than \SI{1}{\kilo\metre\per\second} but we chose to apply it to all of our measured velocity dispersions to replicate the procedure of \cite{Sun2018}.

\begin{figure*}
\includegraphics{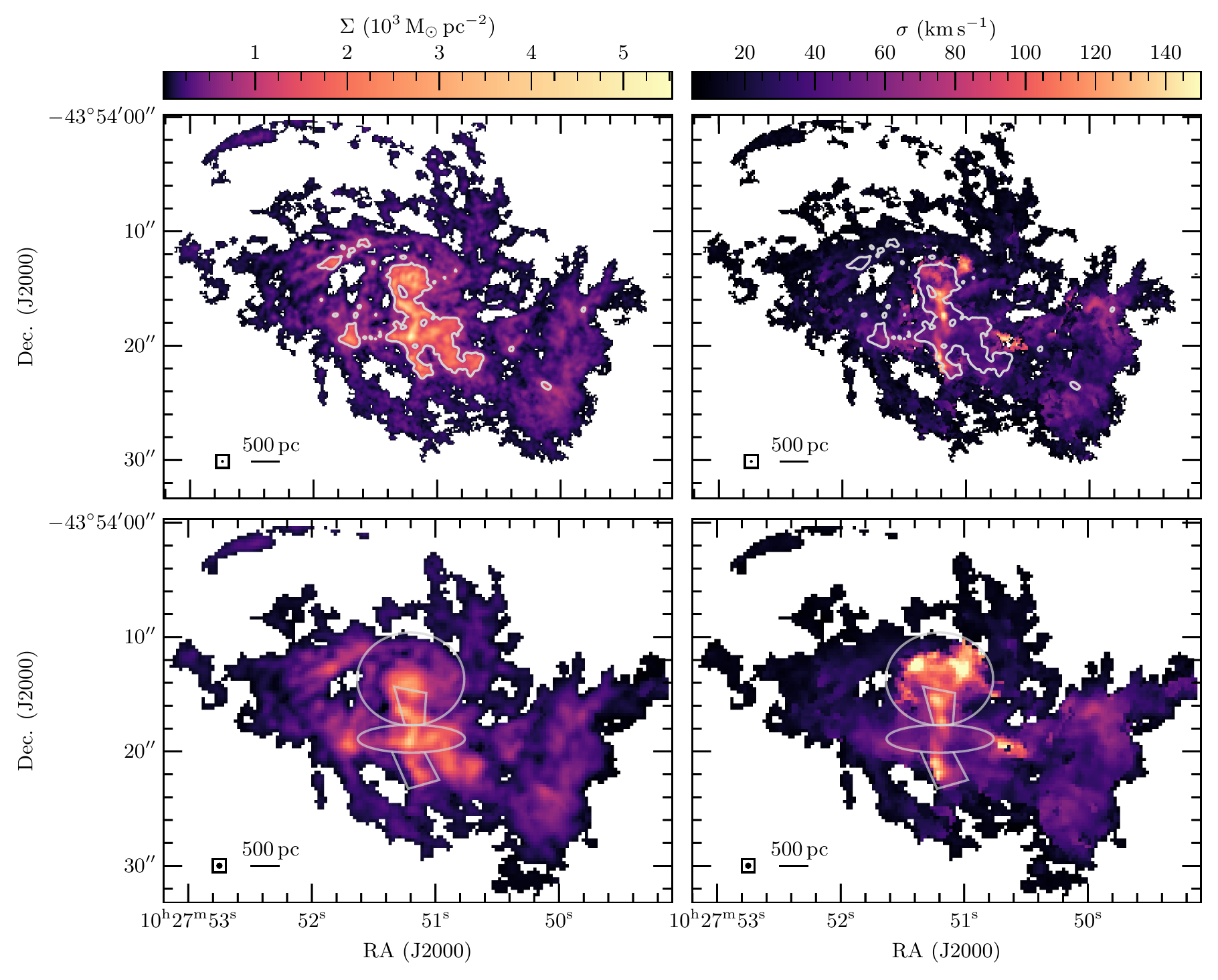}
\caption{Maps of molecular gas mass surface density (left) and velocity dispersion (right), with a synthesized beam \gls{fwhm} of \SI{55}{\parsec} (top) and \SI{120}{\parsec} (bottom). Contours in the top row are at a surface density of \SI{1000}{\solarmass\per\square\parsec}. The same contours are in the left and right frames. Gray ellipses in the bottom row mark the central \SI{1}{\kilo\parsec}, projected to account for the inclination angles of the nuclei. Gray polygons indicate pixels affected by the southern jet, identified in the dispersion maps, that are not included in the distributions of measured quantities shown in subsequent figures. Pixels west of an \gls{ra} of approximately $\num{10}^{\mathrm{h}}\num{27}^{\mathrm{m}}\num{50.3}^{\mathrm{s}}$ are also excluded from our distribution analyses due to complex spectral features and potential interferometric-sidelobe contamination. The beam sizes are represented by the black circles in the bottom-left corners of each frame. The \gls{u/lirg} conversion factor of \SI{1.38}{\solarmass\per\square\parsec}(\si{\kelvin\kilo\metre\per\second})$^{-1}$ was used to convert the integrated intensity maps from \si{\kelvin\kilo\metre\per\second} to \si{\solarmass\per\square\parsec}. Multiplying the surface densities by \num{\sim 4.5} would convert to the Galactic conversion factor of \SI{6.25}{\solarmass\per\square\parsec}(\si{\kelvin\kilo\metre\per\second})$^{-1}$. Both of these conversion factors include a factor of \num{1.36} to account for helium.}
\label{fig:sd_vd_maps}
\end{figure*}

Figure~\ref{fig:sd_vd_maps} shows mass surface density and velocity dispersion maps of \ngc{}, convolved to synthesisized beam sizes of \SIlist{55;120}{\parsec}. Gray contours at the \SI{1000}{\solarmass\per\square\parsec} level are shown in the top row. The bottom row includes gray ellipses marking the inclination-projected central \SI{1}{\kilo\parsec} radii, centred on the positions of the nuclei, and gray polygons marking pixels heavily contaminated by the southern jet, identified in the velocity dispersion maps.

The \gls{fov} shown is centered at \gls{ra} $\num{10}^{\mathrm{h}}\num{27}^{\mathrm{m}}\num{51}^{\mathrm{s}}$ \gls {dec} $\ang[angle-symbol-over-decimal]{-43;54;16}$, and it covers $\num{0.73} \,$arcmin in \gls{ra} and $\num{0.54} \,$arcmin in \gls{dec}. Pixels west of an \gls{ra} of approximately $\num{10}^{\mathrm{h}}\num{27}^{\mathrm{m}}\num{50.3}^{\mathrm{s}}$ are excluded from our analysis. In this region there are elevated velocity dispersions measured coming from two-component spectral profiles. While multicomponent lines near the centre of the system appear to be produced by multiple gas components along the line of sight, we believe some of the two-component profiles in the west are the result of imperfectly cleaned sidelobes of the interferometric synthetic beam. It also appears material from the outskirts of the progenitors are overlapping along the line of sight \citep{Sak2014}, producing spectral profiles that are too complex for the moment-based analysis here. For this reason, we exclude pixels from that part of the maps from the measurements we present.

\section{Results}
\label{results}
In this section, we present mass-weighted Gaussian \gls{kde} distributions of the mass surface densities, velocity dispersions, peak brightness temperatures, virial parameters, and internal turbulent pressures in \ngc{}. The distributions measured in the \gls{phangs} early sample from \cite{Sun2018} are also shown for comparison. Virial parameters and turbulent pressures are calculated in the same way as by \cite{Sun2018} and assuming the diameter of the cloud is equal to the beam \gls{fwhm}.

Distributions include measurements from all pixels containing significant emission and smoothed to \SIlist{55;120}{\parsec}. We also calculated distributions for \SI{80}{\parsec} but do not show them because they always land midway between the distributions from \SIlist{55;120}{\parsec}. A bandwidth of \SI{0.1}{\dex} was used for all distributions except the internal turbulent pressure where a bandwidth of \SI{0.2}{\dex} was used instead. This matches the \gls{kde} calculations of \cite{Sun2018} where the bandwidth for their pressures was misreported as \SI{0.1}{\dex} (J. Sun, private communication).

In the following plots, distributions of pixels from the entire mapped \gls{fov} are shown as solid lines, pixels within a radius of \SI{1}{\kilo\parsec} from the nuclei as dotted lines, and pixels outside the central \SI{1}{\kilo\parsec} radius as dashed lines (referred to here as the non-nuclear pixels and equivalent to the disc pixels in \citealt{Sun2018}). The nuclear peaks in our continuum map were used as the positions of the nuclei.

We then projected the boundaries of the central kiloparsec circles to account for the inclination angles of \SI{30}{\degree} for the northern nucleus \citep{Sak2014} and \SI{75}{\degree} for the southern nucleus. Although \cite{Sak2014} estimate an inclination angle of \SI{80}{\degree} for the southern nucleus, we adopt \SI{75}{\degree} because it better captures the apparent thickness of the southern disc in these data. However, the difference is small as an inclination of \SI{75}{\degree} results in about \num{8} pixels across the minor axis in the \SI{120}{\parsec} maps while \SI{80}{\degree} would result in 5 pixels. The gray ellipses in the bottom row of Figure~\ref{fig:sd_vd_maps} show these central region boundaries. We do not correct the measurements for these inclinations because the disturbed morphology makes it unclear where the boundaries of the inclination regions should be or how to transition between them. There is also the complication of how to deal with overlapping material from the two nuclei. Not including inclination corrections will cause our surface density measurements to overestimate the true values perpendicular to any discs present.

Gray polygons indicate pixels associated with the southern jet and those pixels are excluded from all distributions and total masses used to normalize the \glspl{kde}. Table~\ref{tab:percentiles} presents the \nth{16}, \nth{50}, and \nth{84} percentiles for the different regions of \ngc{}, at each resolution. Comparisons of these percentiles for \ngc{} and the \gls{phangs} sample at \SI{120}{\parsec} resolution are presented in Figures~\ref{fig:1d_sd} to \ref{fig:1d_p_turb}.

\begin{table*}
\caption{Mass-weighted percentiles of all measured and derived quantities from \ngc{}, by region and resolution.}
\begin{threeparttable}
\label{tab:percentiles}
\sisetup{table-number-alignment=center,table-sign-mantissa}
\begin{tabular}{@{}l c S[table-format=4.0] S[table-format=4.0] S[table-format=4.0] S[table-format=2.0] S[table-format=2.0] S[table-format=3.0] S[table-format=1.1] S[table-format=2.1] S[table-format=2.1] S[table-format=1.1] S[table-format=2.1] S[table-format=2.0] S[table-format=3.1] S[table-format=3.0] S[table-format=4.0]@{}}
\hline
Region      & {Beam Size}      & \multicolumn{3}{c}{{$\Sigma$}}                            & \multicolumn{3}{c}{{$\sigma$}}                      & \multicolumn{3}{c}{{$T_{\mathrm{peak}}$}} & \multicolumn{3}{c}{{$\alpha_{\mathrm{vir}}$}} & \multicolumn{3}{c}{{$P_{\mathrm{turb}}$}} \\
            & {(\si{\parsec})} & \multicolumn{3}{c}{{(\si{\solarmass\per\square\parsec})}} & \multicolumn{3}{c}{{(\si{\kilo\metre\per\second})}} & \multicolumn{3}{c}{{(\si{\kelvin)}}}      &      &      &                                 & \multicolumn{3}{c}{{(\SI{1e6}{\kelvin\per\cubic\centi\metre})}} \\
\cmidrule(l{2pt}r{2pt}){3-5}\cmidrule(l{2pt}r{2pt}){6-8}\cmidrule(l{2pt}r{2pt}){9-11}\cmidrule(l{2pt}r{2pt}){12-14}\cmidrule(l{2pt}r{2pt}){15-17}
            &                  & {16} & {50} & {84}                                        & {16} & {50} & {84}                                  & {16} & {50} & {84}                        & {16} & {50} & {84}                            & {16}    & {50}    & {84} \\
\hline
Non-nuclear & 55               &  150 &  390 & 1300                                        &   12 &   25 & 43                                    &  2.7 &  5.1 & 11                          &  5.5 &   11 & 26                              &     2.3 &      26 & 170 \\
            & 80               &  150 &  380 & 1200                                        &   13 &   27 & 45                                    &  2.4 &  4.6 & 10                          &  4.3 &  8.8 & 20                              &     1.8 &      20 & 120 \\
            & 120              &  140 &  370 & 1200                                        &   14 &   28 & 47                                    &  2.2 &  4.2 & 9.5                         &  3.4 &  6.8 & 15                              &     1.3 &      15 & 92  \\
\hline
Nuclei      & 55               &  520 & 1400 & 2500                                        &   27 &   46 & 71                                    &  5.8 &   12 & 19                          &  6.1 &   13 & 28                              &      36 &     270 & 930 \\
Combined    & 80               &  520 & 1400 & 2500                                        &   34 &   55 & 89                                    &  5.4 &   11 & 17                          &  5.6 &   13 & 29                              &      48 &     250 & 970 \\
            & 120              &  500 & 1500 & 2400                                        &   45 &   66 & 100                                   &  5.1 &  9.9 & 16                          &  5.2 &   12 & 34                              &      75 &     210 & 720  \\
\hline
Northern    & 55               &  410 &  870 & 2400                                        &   20 &   42 & 72                                    &  5.1 &   11 & 20                          &  6.2 &   13 & 29                              &      18 &     160 & 820 \\
Nucleus     & 80               &  410 &  870 & 2300                                        &   25 &   62 & 95                                    &  4.8 &  9.7 & 19                          &  6.7 &   18 & 40                              &      18 &     280 & 1000 \\
            & 120              &  390 &  850 & 2200                                        &   44 &   85 & 110                                   &  4.4 &  8.6 & 15                          &  7.4 &   19 & 63                              &      30 &     260 & 800  \\
\hline
Southern    & 55               &  970 & 1700 & 2800                                        &   36 &   49 & 70                                    &  6.5 &   13 & 18                          &  5.9 &   12 & 24                              &     160 &     330 & 1200 \\
Nucleus     & 80               &  980 & 1700 & 2600                                        &   39 &   51 & 73                                    &  6.3 &   12 & 17                          &  4.7 &  9.7 & 19                              &     130 &     250 & 940 \\
            & 120              & 1100 & 1700 & 2500                                        &   45 &   54 & 76                                    &  6.4 &   12 & 16                          &  4.1 &  7.6 & 13                              &      98 &     190 & 620  \\
\hline
\end{tabular}
\end{threeparttable}
\end{table*}

\subsection{Mass surface density}
\begin{figure*}
\includegraphics{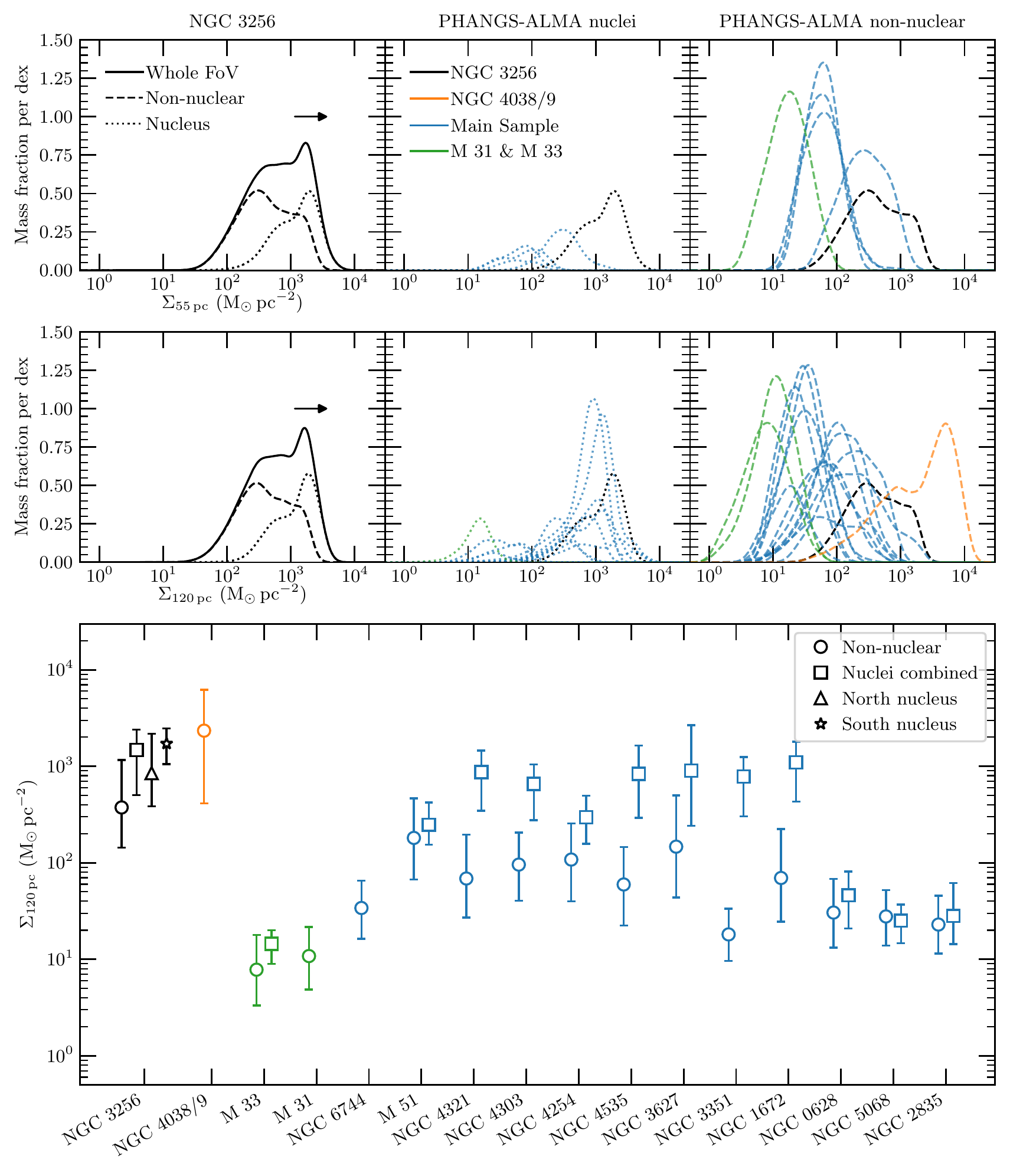}
\caption{Molecular gas mass surface densities measured in \ngc{} and the \protect\cite{Sun2018} \gls{phangs} sample. Measurements for \ngc{} are made in maps with beam \gls{fwhm} of \SIlist{55;120}{\parsec} and pixels half those lengths on a side. \gls{phangs} distributions come from maps with beam \gls{fwhm} of \SIlist{45;120}{\parsec} and pixels half those lengths on a side. \textbf{\emph{Top and middle rows:}} Solid lines are pixel distributions from the entire \gls{fov}, dashed lines are non-nuclear region pixels, and dotted lines are nuclear region pixels. Arrows show how values would shift if we adopt the Milky Way \gls{co} conversion factor of \SI{6.25}{\solarmass\per\square\parsec}(\si{\kelvin\kilo\metre\per\second})$^{-1}$ (used for all \gls{phangs} galaxies) instead of the \gls{lirg} conversion factor of \SI{1.38}{\solarmass\per\square\parsec}(\si{\kelvin\kilo\metre\per\second})$^{-1}$. \gls{phangs} galaxies are separated into nuclear and disc regions in the centre and right panels, respectively, with the corresponding curves from \ngc{} reproduced from the left panels. \textbf{\emph{Bottom row:}} Mass-weighted medians shown as symbols and ranges encompassing the inner \num{68} per cent of the distributions shown as error bars for \ngc{} and the \protect\cite{Sun2018} sample. All percentiles are calculated from the \SI{120}{\parsec} maps.}
\label{fig:1d_sd}
\end{figure*}

At all resolutions, \ngc{} exhibits surface density distributions with a maximum centred near \SI{2000}{\solarmass\per\square\parsec} and a shoulder centred near \SI{600}{\solarmass\per\square\parsec} (see Figure~\ref{fig:1d_sd}). We measure that at least \num{85} per cent of the area out to \SI{2}{\kilo\parsec} from the midpoint between the two nuclei has $\Sigma \gtrsim \text{\SI{100}{\solarmass\per\square\parsec}}$, independent of resolution.

The mass in the higher surface density peak of the distribution is dominated by pixels in the nuclear regions, but there is a significant contribution from pixels outside the central kiloparsec as well. Contours in the top row of  Figure~\ref{fig:sd_vd_maps} show the pixels with $\Sigma>\SI{1000}{\solarmass\per\square\parsec}$. The majority of these pixels lie in the nuclear and southern jet regions, though not all nuclear pixels are at these high surface densities.

The lower surface density shoulder of the pixel distribution is a roughly equal mix of nuclear and non-nuclear pixels, but switches to being dominated by the non-nuclear pixels around \SI{600}{\solarmass\per\square\parsec}. The central high density region is better described in the northern nucleus of \ngc{} by a radius of \SI{250}{\parsec} than the \SI{1}{\kilo\parsec} used for the \gls{phangs} galaxies. Considering the extra-nuclear high surface density lines of sight and low surface density pixels within the central kiloparsec, a careful decomposition of the gas emission using spatial and spectral dimensions will be crucial to fully understand the complex morphology of this merger. The pixel distributions throughout this section provide bulk measurements of the gas properties, but evaluation of those properties ultimately needs to be done on a sightline-by-sightline basis.

Comparing the mass surface densities measured in \ngc{} to \gls{phangs} in the bottom row of Figure~\ref{fig:1d_sd}, \ngc{} occupies the high end of the distributions observed in the \gls{phangs} sample. The nuclei of most \gls{phangs} galaxies are consistent with the nuclei and upper range of the non-nuclear distribution in \ngc{}. Most disc distributions from \gls{phangs} overlap with the lower half of the non-nuclear distribution in \ngc{}.

\subsection{Velocity dispersion}\label{dispersion_results}
\begin{figure*}
\includegraphics{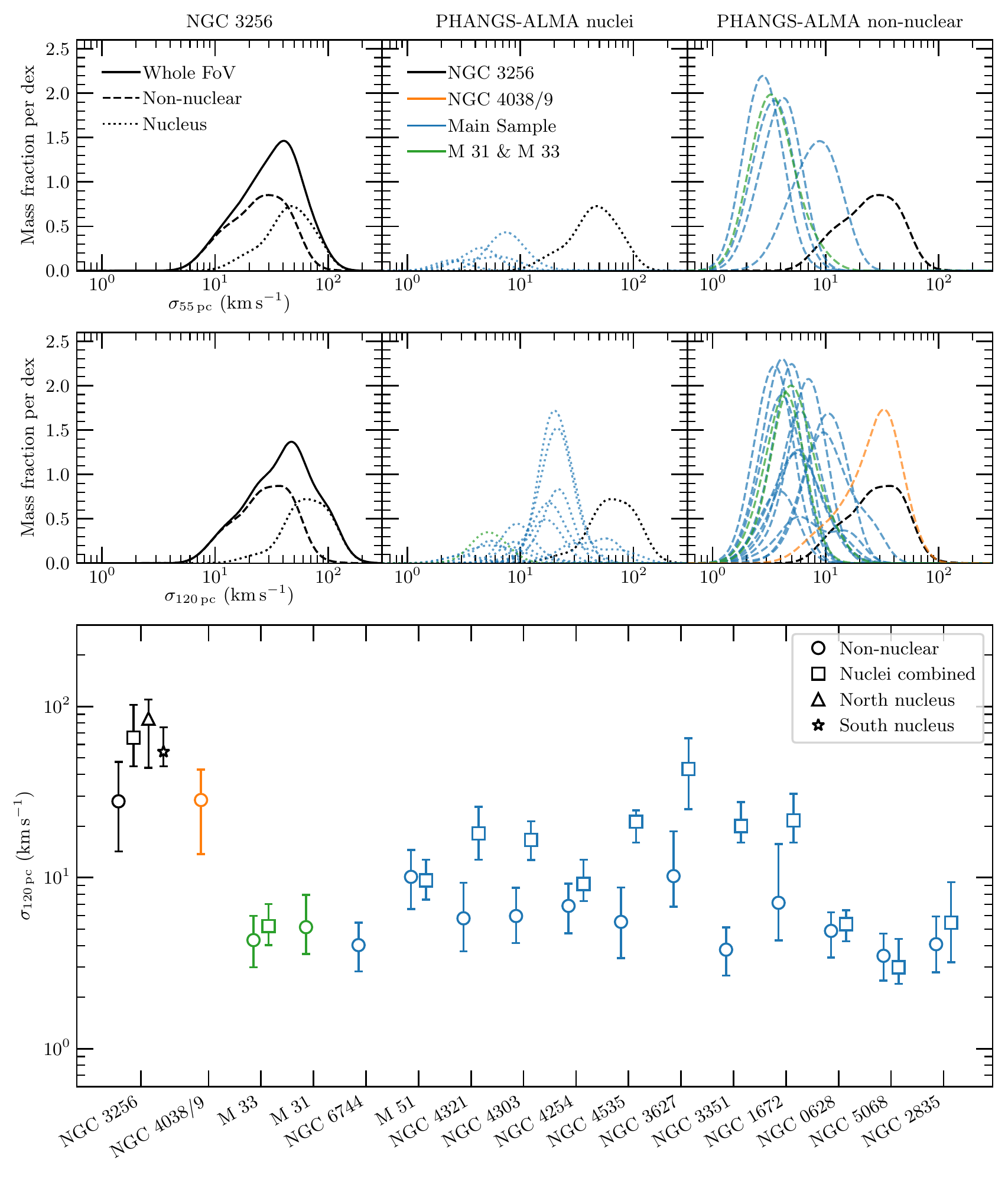}
\caption{Same as Figure~\ref{fig:1d_sd} except for distributions of velocity dispersion measurements.}
\label{fig:1d_vd}
\end{figure*}

Velocity dispersions in \ngc{} from all pixels and resolutions show a broad distribution in Figure~\ref{fig:1d_vd}, ranging from \SIrange{10}{200}{\kilo\metre\per\second} and with a single peak near \SI{50}{\kilo\metre\per\second}. The distribution does broaden towards higher dispersions going from \SIrange{55}{120}{\parsec} resolution. The non-nuclear distribution peaks near \SI{30}{\kilo\metre\per\second} (independent of resolution), and the nuclear distribution peak increases from near \SI{50}{\kilo\metre\per\second} at \SI{55}{\parsec} to \SI{70}{\kilo\metre\per\second} at \SI{120}{\parsec}. Separating the nuclei in Table~\ref{tab:percentiles} shows that while both nuclei shift to higher dispersions at lower resolution, the northern nucleus has a much stronger trend with resolution. The trend in the northern nucleus contains significant contamination from the outflow, driven by the northern nucleus, that becomes worse as the resolution is made coarser. This is due to improved surface brightness sensitivity at lower resolution increasing the detection of very wide spectral wings from the outflow. The high dispersion blobs in the \SI{120}{\parsec} resolution map that are mostly absent in the \SI{55}{\parsec} map are sightlines that are most affected by the outflow contribution.

Figure~\ref{fig:1d_vd} shows that the velocity dispersions for \ngc{} are significantly higher than most of the \gls{phangs} sample. Almost no dispersions in the \gls{phangs} sample are as high as in the nuclei of \ngc{}, except for NGC~3627 discussed in Section~\ref{3627_compare}. The nuclei of some \gls{phangs} galaxies overlap with the lower half of the \ngc{} non-nuclear distribution.

\subsection{Peak brightness temperature}
\begin{figure*}
\includegraphics{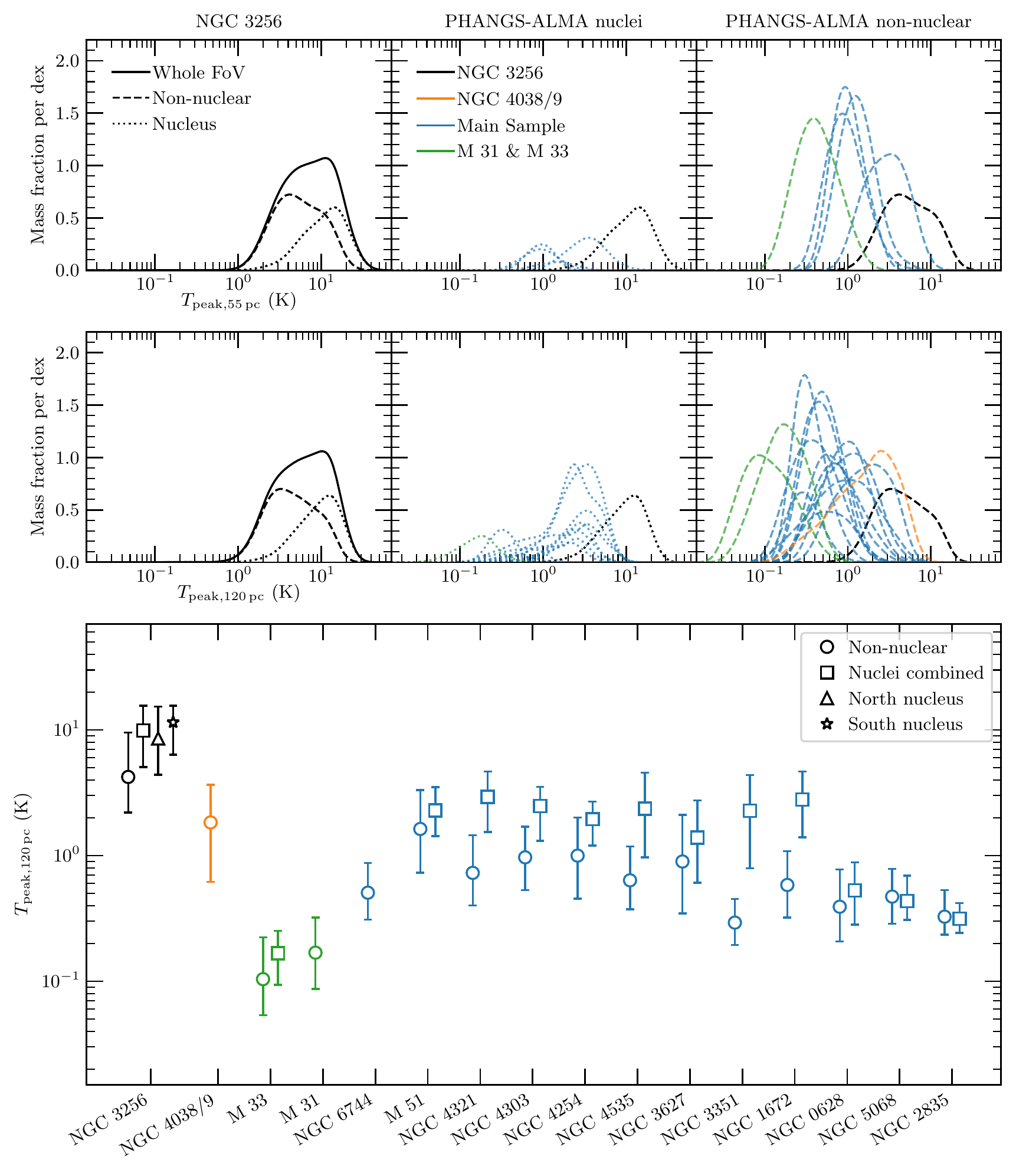}
\caption{Same as Figure~\ref{fig:1d_sd} except for distributions of peak brightness temperature measurements.}
\label{fig:1d_peak_temp}
\end{figure*}

The peak brightness temperature distributions from all pixels in \ngc{} are centred near \SI{7}{\kelvin}, range from \SIrange{1}{40}{\kelvin}, are slightly skewed to higher temperatures, and do not change significantly with resolution (see Figure~\ref{fig:1d_peak_temp}). The non-nuclear distribution peaks at \SI{\sim 3}{\kelvin} and the combined nuclei distribution peaks at \SI{\sim 15}{\kelvin}, but the inner \num{68} per cent of both distributions overlap significantly (\SIrange{3}{11}{\kelvin} for the non-nuclear and \SIrange{6}{19}{\kelvin} for the nuclei).

While the nuclear and non-nuclear distributions in \ngc{} overlap considerably, there is still a difference of about two in the median brightness temperatures between them. Assuming the sizes of molecular structures are not significantly different between those regions, this difference would indicate gas kinetic temperatures that are at most two times higher in the nuclei.

Similar to the velocity dispersions, the brightness temperatures in \ngc{} reach higher values than all of the galaxies in the \gls{phangs} sample. Figure~\ref{fig:1d_peak_temp} shows that most of the \gls{phangs} centres and the disc of M51 are consistent with only the non-nuclear distribution from \ngc{}.

\subsection{Virial parameter}\label{vir_results}
\begin{figure*}
\includegraphics{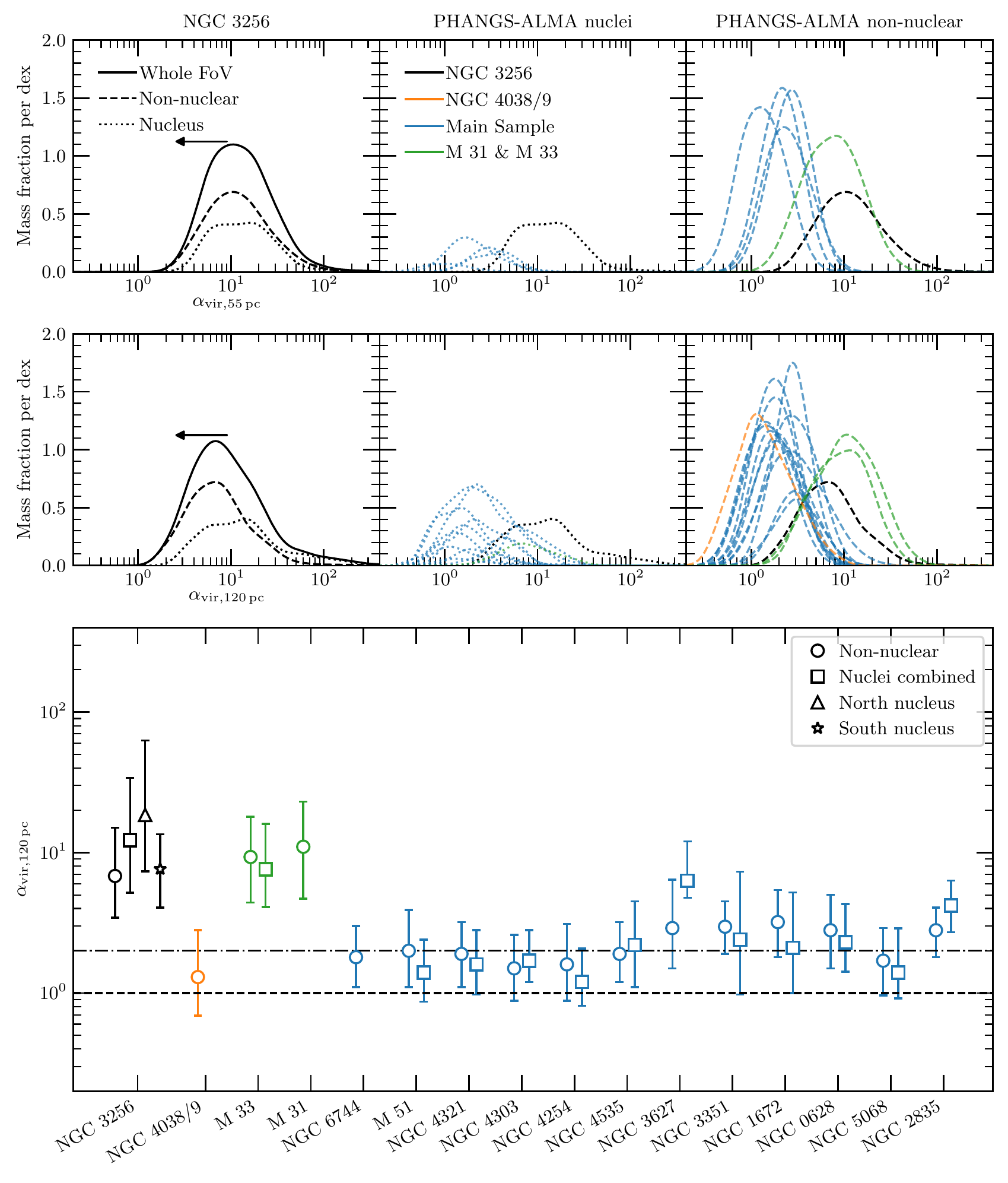}
\caption{Same as Figure~\ref{fig:1d_sd} except for distributions of virial parameter measurements. The horizontal dashed line shows $\alpha_{\mathrm{vir}} = 1$ (virial equilibrium) and the dash-dotted line shows $\alpha_{\mathrm{vir}} = 2$ (approximately bound or collapsing). The \gls{u/lirg} conversion factor of \SI{1.38}{\solarmass\per\square\parsec}(\si{\kelvin\kilo\metre\per\second})$^{-1}$ was used to calculate the surface densities for \ngc{}. Dividing the virial parameters of \ngc{} by \num{\sim 4.5} would convert to the Galactic conversion factor of \SI{6.25}{\solarmass\per\square\parsec}(\si{\kelvin\kilo\metre\per\second})$^{-1}$ (used for all \gls{phangs} galaxies).}
\label{fig:1d_alpha_vir}
\end{figure*}

Figure~\ref{fig:1d_alpha_vir} shows the virial parameter distributions for all pixels in \ngc{} peak around \num{10} but with long tails to near \num{1} and above \num{100}. Changing the resolution slightly broadens the distributions when going from \SIrange{55}{120}{\parsec}. This occurs because at \SI{55}{\parsec} the nuclear and non-nuclear distributions are very similar (except for more mass present in the non-nuclear pixels). At lower resolution, the non-nuclear pixels shift towards lower virial parameters, peaking around \num{6} at \SI{120}{\parsec}, while the peak of the nuclear pixel distribution does not move. Table~\ref{tab:percentiles} shows these changes are even more complex since the northern nucleus exhibits higher virial parameters at lower resolution but the southern nucleus shows the opposite trend. Non-nuclear virial parameters become slightly lower at lower resolution, but most of the mass has overlapping distributions between resolutions. In the simple framework of balancing internal kinetic energy with self-gravity, the gas at these scales is not bound by gravity alone. It is worth highlighting here that changing from the \gls{u/lirg} to the Galactic conversion factor would result in the distributions being centered around two and would change the gas from being clearly unbound to marginally bound. Further analysis will have to wait for a better constrained \gls{co} conversion factor, possibly through a cloud decomposition of these data.

\ngc{} appears quite different from most of the \gls{phangs} sample in virial parameter, as shown in Figure~\ref{fig:1d_alpha_vir}. Most of the \gls{phangs} distributions hover around $\alpha_{\mathrm{vir}}$ of about one or two. Roughly half of the galaxies in the \gls{phangs} sample sit well below the distributions measured in \ngc{}. Interestingly, M31 and M33 are the most consistent with the distributions from \ngc{} which may be due to properties unique to the \gls{ism} regime in those galaxies as well as observational effects \citep[see Sections~5.2.4 and 5.2.5 from][]{Sun2018}. A combination of reduced beam-filling factors at low surface densities, missing self-gravity from an underestimated \gls{co} conversion factor (also at low surface densities), and more dominant external pressures from larger fractions of atomic gas may be driving the higher virial parameters seen in M31 and M33.

\subsection{Internal turbulent pressure}
\begin{figure*}
\includegraphics{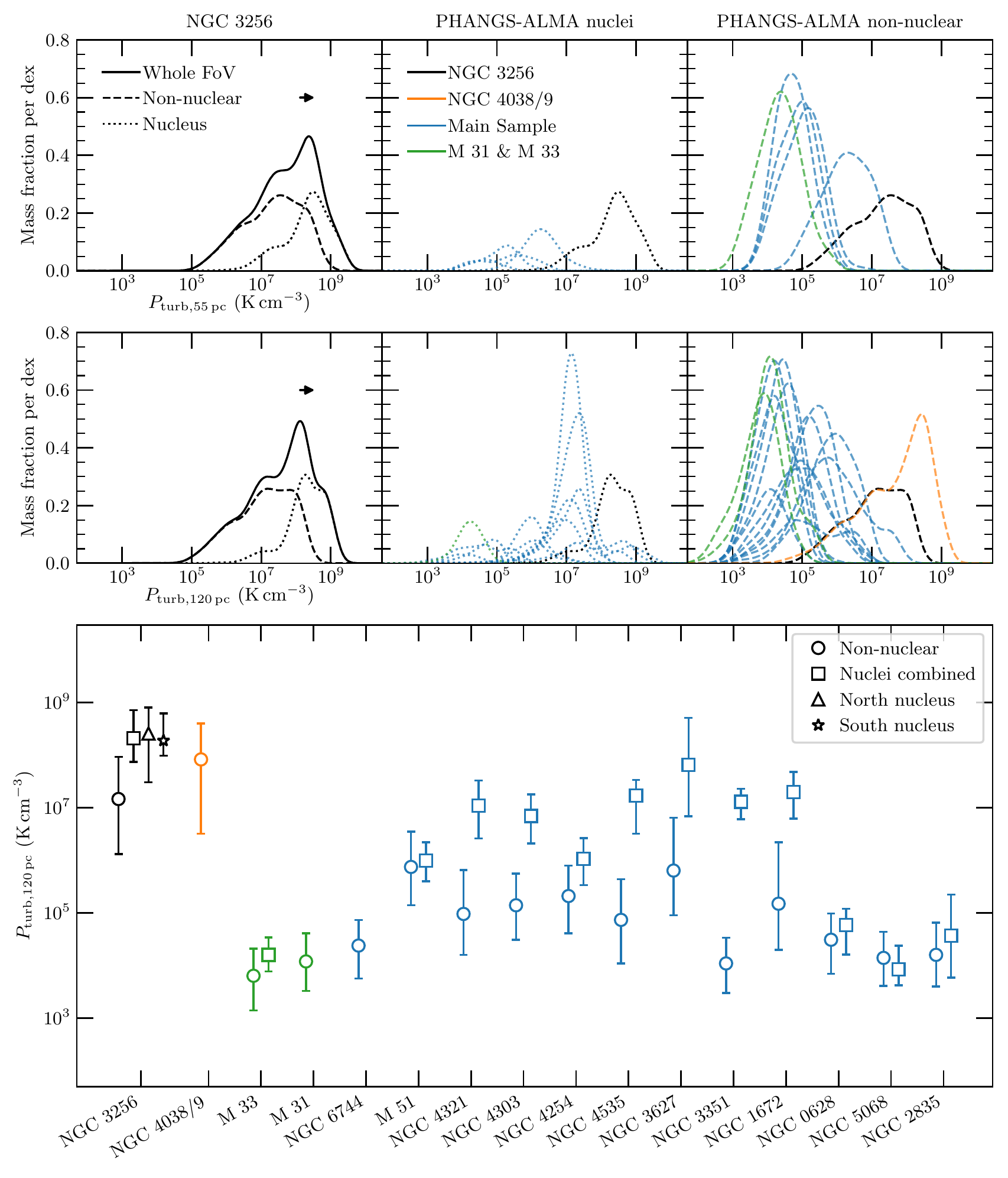}
\caption{Same as Figure~\ref{fig:1d_sd} except for distributions of internal turbulent pressure measurements. The \gls{u/lirg} conversion factor of \SI{1.38}{\solarmass\per\square\parsec}(\si{\kelvin\kilo\metre\per\second})$^{-1}$ was used to calculate the surface densities used here for \ngc{}. Multiplying the pressures by \num{\sim 4.5} would convert to the Galactic conversion factor of \SI{6.25}{\solarmass\per\square\parsec}(\si{\kelvin\kilo\metre\per\second})$^{-1}$ (used for all \gls{phangs} galaxies).}
\label{fig:1d_p_turb}
\end{figure*}

Internal turbulent pressure measured in all pixels exhibits the most complex distribution, with up to four peaks spanning a range from \SIrange{1e5}{1e10}{\kelvin\per\cubic\centi\metre} (see Figure~\ref{fig:1d_p_turb}). It appears most of the mass at the highest pressures originates from the nuclear regions, and that the nuclear regions have almost a single narrowly peaked distribution around \SI{5e8}{\kelvin\per\cubic\centi\metre}. The additional nuclear component near \SI{1e7}{\kelvin\per\cubic\centi\metre} comes from pixels with surface densities below \SI{1000}{\solarmass\per\square\parsec} (i.e. pixels not within the contours in Figure~\ref{fig:sd_vd_maps}) but that still lie within \SI{1}{\kilo\parsec} of the centre of the northern nucleus. While close to the northern nucleus in projection, these pixels have pressures consistent with the distribution from the non-nuclear pixels, indicating the gas properties are only at their most extreme very close to the centres of the nuclei. The southern nuclear region does not include such low pressures.

The comparison to the \gls{phangs} sample in Figure~\ref{fig:1d_p_turb} is quite similar to that for the velocity dispersion. The main difference is that the nuclear pressure distributions from \ngc{} are not as extreme as the dispersions. This results in some of the \gls{phangs} main sample nuclei being consistent with the nuclei of \ngc{}. There is still a large fraction of discs and even whole galaxies from \gls{phangs} that have pressures two to three orders of magnitude lower than \ngc{}. The effect of the conversion factor choice is relatively small for the measured pressures given the width of the distributions. If the assumptions used to derive the turbulent pressure hold in \ngc{} it contains some of the highest pressures in this sample of galaxies.

\subsection{Two-dimensional view}
\begin{figure*}
\includegraphics{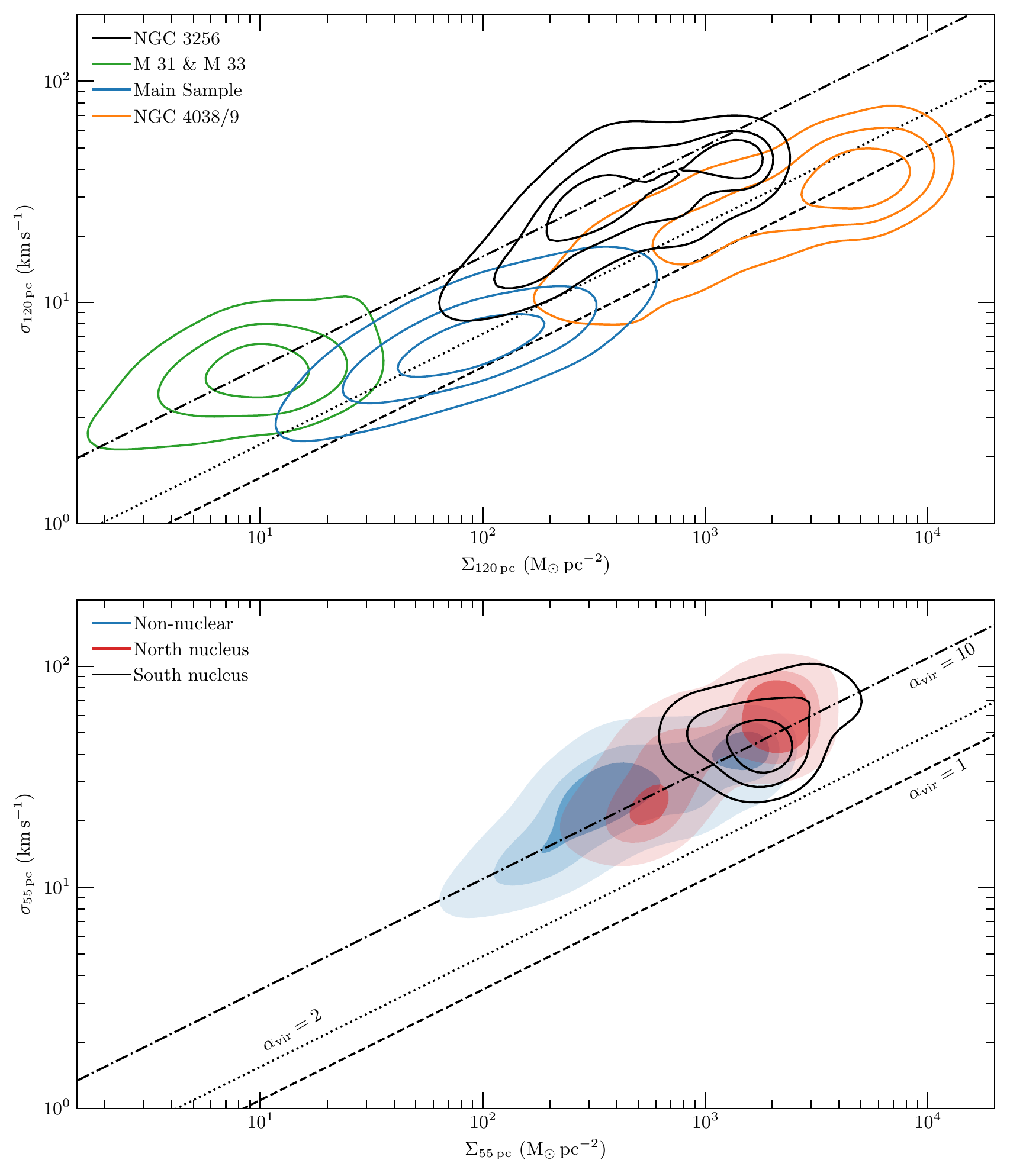}
\caption{Velocity dispersion vs. mass surface density with contours enclosing \numlist{20;50;80} per cent of the total mass. Lines of constant virial parameter are shown to emphasize the difference in dynamical state between the \gls{phangs} main sample and \ngc{}. All \gls{phangs} galaxies use the Galactic \gls{co} conversion factor of \SI{6.25}{\solarmass\per\square\parsec}(\si{\kelvin\kilo\metre\per\second})$^{-1}$ while \ngc{} uses the \gls{u/lirg} of \SI{1.38}{\solarmass\per\square\parsec}(\si{\kelvin\kilo\metre\per\second})$^{-1}$. \textbf{\emph{Top:}} Distributions from non-nuclear regions only, measured in \SI{120}{\parsec} resolution maps. \textbf{\emph{Bottom:}} All regions observed in \ngc{} at \SI{55}{\parsec} resolution.}
\label{fig:2d_KDE_vd_vs_sd}
\end{figure*}

Two-dimensional Gaussian \glspl{kde} of non-nuclear velocity dispersion versus surface density are shown in the top panel of Figure~\ref{fig:2d_KDE_vd_vs_sd}. A bandwidth of \num{0.1} dex was used, and contours enclose \numlist{20;50;80} per cent of the mass.

\ngc{} is offset from the trend in the \gls{phangs} main sample and overlap region of NGC~4038/9. Most of this offset would be removed by switching \ngc{} to the Galactic \gls{co} conversion factor. Figure~\ref{fig:2d_KDE_vd_vs_sd} makes it obvious that when using the \gls{u/lirg} conversion factor there is significant overlap in non-nuclear surface densities in \ngc{} and the \gls{phangs} main sample. If the surface densities for \ngc{} are correct then the main driver pushing it off the \gls{phangs} trend is its high velocity dispersions.

The bottom panel of Figure~\ref{fig:2d_KDE_vd_vs_sd} shows the same two-dimensional \gls{kde} but for \ngc{} alone, at \SI{55}{\parsec} resolution. Separate distributions for the non-nuclear, northern, and southern nucleus pixels are shown. The gap between peaks in the non-nuclear distribution is wider at higher resolution, driven mainly by a change in the surface densities. A similar bimodality is present in the northern nucleus, but the peaks are slightly shifted to higher surface densities than the non-nuclear peaks. The southern nucleus exhibits a unimodal distribution with its peak quite similar to the higher non-nuclear peak.

The bimodal distribution in the non-nuclear pixels appears to originate from measurements of truly non-nuclear pixels and some pixels that contain a contribution from the nuclear jet and/or outflow activity. Since the higher peak in the non-nuclear distribution is slightly closer to the peak in the southern nucleus distribution, it may be that the southern nucleus is the main contaminant to the non-nuclear pixels. This is reasonable since the jet from the southern nucleus may not be perfectly excluded from the non-nuclear pixels. It is also more likely the central \SI{1}{\kilo\parsec} boundary is not as accurate as for the northern nucleus due to the nearly edge-on orientation of the southern nucleus.

Bimodality in the northern nucleus distribution again points to the central gas structure being made up of a central region of high surface density with radius \SI{\sim 250}{\parsec} surrounded by more non-nuclear-like gas from \SI{250}{\parsec} outward.

\subsection{Two detailed comparison studies}
Here we provide a more detailed comparison of two galaxies with \ngc{}, as they each present interesting cases. NGC~4038/9 is an important system for comparison because it is the only other merger analyzed at the same resolution and in the same manner. It is the nearest merger at \SI{22}{\mega\parsec} \citep{Sch2008} and contains \SI{\sim 2e10}{\solarmass} in molecular gas \citep{Gao2001}. Star formation rates vary throughout the system, with the overlap region analyzed by \cite{Sun2018} \citep[data originally presented by][]{Whi2014} undergoing vigorous star formation \citep[\SI{> 4}{\solarmass\per\year},][]{Bra2009, Kla2010, Bem2019}. Simulations place it at an intermediate merger stage, \SI{\sim 40}{\mega\year} after its second pass \citep{Kar2010}. There does not appear to be any conclusive evidence of currently active nuclear outflows in the system.

We also examine NGC~3627 because it has extreme properties similar to \ngc{} that often set it apart from the rest of the main \gls{phangs} sample. It is also a useful galaxy for comparison because it appears to follow the trends of extreme surface densities and dispersions relative to unbarred spiral galaxies shown by \cite{Sun2020b} in the full \gls{phangs} of \num{70} galaxies. It is at a distance of \SI{\sim 11}{\mega\parsec} \citep{Lee2013}, is part of the Leo Triplet, shows evidence for LINER/Seyfert 2 nuclear activity \citep{Pen1998}, has a stellar mass of \SI{\sim 1e11}{\solarmass} \citep{Kar2014}, and molecular gas mass of \SI{\sim 1e10}{\solarmass} \citep{Law2018}. Optical images show a bar and two asymmetric spiral arms \citep{Pta2006} likely originating from interactions with NGC~3628 \citep{Rot1978, Soi2001}. These morphological features also stand out in \gls{co} observations \citep{Law2018, Sun2018}. There is no evidence for currently active nuclear outflows in this system either.

\subsubsection{NGC 4038/9}
The surface density distribution of NGC~4038/9 is the only one to significantly exceed all measurements of surface densities from \ngc{}, as shown in Figure~\ref{fig:1d_sd}. About \num{50} per cent of the gas mass in the overlap region lies at surface densities above \ngc{}. Both the non-nuclear pixels and the nuclei of \ngc{} are consistent with the lower half of the distribution from NGC~4038/9. About half the mass in the non-nuclear pixels of \ngc{} lies at surface densities below the NGC~4038/9 distribution. It is important to note that the Galactic conversion factor was used to calculate the mass surface densities for NGC~4038/9, as was done by \cite{Sun2018}. In Section~\ref{caveats}, we discuss the applicability of the Galactic conversion factor to the overlap region of NGC~4038/9, and how the comparison to \ngc{} changes if the \gls{u/lirg} conversion factor was used instead.

Figure~\ref{fig:1d_vd} shows that medians and widths of the velocity dispersion distributions in NGC~4038/9 and the non-nuclear pixels of \ngc{} are very similar. The distributions also show similar shapes, with a sharper drop at high dispersions and a longer tail at low dispersions. However, the majority of mass in the nuclear pixels of \ngc{} has velocity dispersions higher than the NGC~4038/9 distribution.

For peak brightness temperatures in Figure~\ref{fig:1d_peak_temp}, most of the gas measured in \ngc{} is at higher values than NGC~4038/9. The lower half of the non-nuclear distribution in \ngc{} overlaps with the upper half of the NGC~4038/9 distribution. Almost all of the mass in the nuclei of \ngc{} has peak brightness temperatures above NGC~4038/9. This difference could be the result of higher molecular gas kinetic temperatures in \ngc{} compared to the overlap region. It is reasonable to expect the gas to be warmer in \ngc{} when its star formation rate surface densities range from \SIrange{1.3}{3.5}{\solarmass\per\year\per\square\kilo\parsec} on \SI{512}{\parsec} scales in the non-nuclear regions \citep{Wil2019} compared to \SIrange{0.21}{0.84}{\solarmass\per\year\per\square\kilo\parsec} in the overlap region on kiloparsec scales \citep{Bem2019}. Additionally, a $^{12}$\gls{co}/$^{13}$\gls{co} J=\num{1}--\num{0} ratio towards \ngc{} of \num{35} \citep{Aal1991a} and of \num{13} towards the overlap region of NGC~4038/9 \citep{Aal1991b} may be indicative of higher gas temperatures in \ngc{}. However, while the difference in this line ratio likely indicates a difference in the gas properties of these two systems, additional transitions must be observed and modeled to determine if differing gas temperatures are truly present. For example, \cite{Zhu2003} included \num{2}--\num{1} and \num{3}--\num{2} ratios for some positions in the overlap region and still arrived at large velocity gradient solutions with temperatures of \SI{\sim 30}{\kelvin} or \SI{120}{\kelvin}. The ratio can also be complicated by the different merger stages where late-stage mergers likely have more $^{12}$\gls{co} produced via massive  star formation processes \citep[e.g.][]{Sli2017, Bro2019}.

The combination of higher surface densities but similar velocity dispersions results in virial parameters in NGC~4038/9 being significantly below all regions in \ngc{} (see Figure~\ref{fig:1d_alpha_vir}). NGC~4038/9 shows a very similar distribution to most of the galaxies in the \gls{phangs} main sample. This can be seen in Figure~\ref{fig:2d_KDE_vd_vs_sd} where NGC~4038/9 extends the trend seen in the discs of the main sample. The internal turbulent pressures calculated for NGC~4038/9 are consistent with all but the lowest portion of the \ngc{} distributions, as shown in Figure~\ref{fig:1d_p_turb}. Such similar internal pressures between the two mergers may be the result of similar driving mechanisms of turbulence, but their differing virial parameters appear to indicate dynamically important differences in the state of their gas density enhancements. Caveats related to the calculation of virial parameter and turbulent pressure, which also apply to NGC~4038/9, are discussed in Section~\ref{discussion}.

\subsubsection{NGC 3627}\label{3627_compare}
Figure~\ref{fig:1d_sd} shows that the disc of NGC~3627 has some of the highest surface densities among the \gls{phangs} main sample, only matched by M51 and the centre of NGC~4254. The upper half of the disc distribution overlaps with the non-nuclear and northern nucleus distribution of \ngc{}. The centre of NGC~3627 has the highest measured surface densities in the main sample, and even slightly exceeding the highest values from \ngc{}. Only the overlap region of NGC~4038/9 has higher surface densities. This is consistent with the comparison by \cite{Sun2020b} of the centres of \num{43} barred spiral galaxies having a \num{\sim 20} times higher mass-weigthed median surface density than pixels outside the central regions of \num{13} unbarred spiral galaxies. The centre of NGC~3627 also exhibits the broadest distribution of all of the centres from the main sample shown in Figure~\ref{fig:1d_sd} as well as the nuclei of \ngc{}. Only the distribution from the overlap region of NGC~4038/9 is wider.

Velocity dispersions in the disc of NGC~3627 lie above the majority of the measurements in the main sample in Figure~\ref{fig:1d_vd}, but they only overlap with the lower \num{\sim 25} percent of the non-nuclear distribution from \ngc{}. However, the central distribution in NGC~3627 is consistent with both the upper half of the non-nuclear distribution and the lower half of the nuclei in \ngc{}. This is again consistent with the full \gls{phangs} sample where \cite{Sun2020b} measure a \num{\sim 5} times higher mass-weighted velocity dispersion in their barred compared to unbarred galaxies, and again the width of the central distribution in NGC~3627 is one of the broadest compared to all centres from \gls{phangs} shown in Figure~\ref{fig:1d_vd}.

Despite relatively wide distributions of peak brightness temperatures in NGC~3627 (see Figure~\ref{fig:1d_peak_temp}), both the disc and centre are consistent with most of the main sample. Brightness temperature distributions from NGC~3627 overlap with roughly the lower half of temperatures measured in the non-nuclear pixels of \ngc{}. The surface densities and velocity dispersions are similar to the mergers analyzed here, and may be due to recent interactions with NGC~3628. However, the peak brightness temperatures of NGC~3627 are more in line with its relatively modest star formation rate of \SI{\sim 5}{\solarmass\per\year} \citep{Cal2015}, or surface densities on about \SI{300}{\parsec} scales reaching up to \SI{\sim 0.3}{\solarmass\per\year\per\square\kilo\parsec} \citep{Wat2011}. \cite{Cor2018} report $^{12}$CO/$^{13}$CO J=\num{1}--\num{0} ratios around \num{10} across NGC~3627, making it more similar to the overlap region in NGC~4038/9 than \ngc{}. This could be another case of kinetic temperatures being higher in \ngc{}, which is particularly interesting given how similar the other gas properties are between it and NGC~3627.

The centre of NGC~3627 stands out as having the highest virial parameters in the \gls{phangs} main sample in Figure~\ref{fig:1d_alpha_vir}. It overlaps with the upper half of the non-nuclear distribution in \ngc{}, and is also consistent with both nuclei. The disc also exhibits quite high virial parameters for the main sample, overlapping with the non-nuclear and southern nucleus distributions from \ngc{}. This consistency is driven by the similar surface densities and velocity dispersions measured in NGC~3627 and \ngc{}. Figure~\ref{fig:1d_p_turb} shows similar results for the internal turbulent pressures. \cite{Sun2018} show the dispersions versus surface densities for just NGC~3627, split into disc and central regions, exhibiting a trend offset from the main sample like \ngc{}.

\section{Discussion}
\label{discussion}
\subsection{\glsentrytext{ism} structure}
\cite{Sun2018} measure increasing median velocity dispersions with increasing beam \gls{fwhm} (between \SIlist{0.07;0.18}{\dex} in the discs and \SIlist{0.03;0.1}{\dex} in the centres). This is indicative of the structured nature of the turbulent \gls{ism} because turbulent motions are expected to decrease at smaller scales \citep[as in the size-linewidth relation;][]{Lar1981}. A correlation between velocity dispersion and beam size is also measured in \ngc{},  but is very weak in all regions except the northern nucleus. The median dispersions in the non-nuclear region pixels increase by \SI{0.05}{\dex}, in the northern nucleus by \SI{0.31}{\dex} (see Section~\ref{dispersion_results} for a discussion of how increased sensitivity to the nuclear outflow at lower resolution is likely the source of this trend), and in the southern nucleus by \SI{0.04}{\dex}. So despite more than a factor of two change in physical beam size, there does not appear to be evidence for the turbulent size-linewidth relation at these scales.

Another signature of the structured \gls{ism} in the \gls{phangs} sample is in the decreasing median surface densities with increasing beam size (between \SIlist{0.17;0.39}{\dex} in the discs and \SIlist{0.08;0.45}{\dex} in the centres). This is caused by decreasing beam dilution as the resolution approaches the size of the clouds making up the \gls{ism}. In contrast, median surface densities in \ngc{} only decrease by \SI{0.02}{\dex}, \SI{0.01}{\dex}, and not at all in the non-nuclear, northern, and southern nucleus, respectively. If the molecular gas is smooth across the physical scales analyzed here, and star-forming overdensities primarily exist on scales smaller than \SI{55}{\parsec}, then we would expect no trend in surface brightness with resolution.

We can also use our measurements of the peak brightness temperatures to probe the structure of the gas in \ngc{}. Given the optically thick nature of low-J transitions of \gls{co}, the brightness temperature approximately traces the kinetic temperature of the molecular gas. Taking the beam filling factor into account means the brightness temperature actually represents a lower limit for the kinetic temperature. If the emitting clouds are smaller than our \SI{55}{\parsec} beam then the change in filling factor between \SIlist{120;55}{\parsec} would be expected to result in a change of \SI{0.68}{\dex}, the ratio of beam areas. This scenario does not fit because we see only a \SI{0.08}{\dex} increase in the median brightness temperature in the \SI{55}{\parsec} measurements compared to \SI{120}{\parsec}.

If we assume the clouds are spherical we can instead solve for the diameter that would give the measured change in brightness temperature between \SIlist{55;120}{\parsec}. Doing so we find the structures need to be \SI{\sim 110}{\parsec} across. Since we should be able to resolve this size in our \SI{55}{\parsec} map, this size is not consistent with the constant surface densities and dispersions measured as a function of beam size. This scenario would also not explain why the brightness temperature increases marginally between \SIlist{80;55}{\parsec} and by roughly the same factor as between \SIlist{120;80}{\parsec}.

The median temperature change can be explained if both the filling factor and the kinetic temperature of the gas are allowed to change between resolutions. However, if we assume the emitting sources are smaller than \SI{55}{\parsec} and adopt the geometric change in the filling factor expected between our largest and smallest beam, then the kinetic temperatures would have to decrease by a factor of four as the beam \gls{fwhm} decreases, which seems excessive.

One option not ruled out is that the bulk of the molecular gas is in a smooth medium that largely fills the volume but does not break up into higher density clouds, even at \SI{55}{\parsec} scales. In this scenario, the similarity of brightness temperature distributions across resolutions arises for the same reason we see no trend in surface density with resolution. So the lack of any strong trends with resolution in velocity dispersion, surface density, and brightness temperature indicates the molecular \gls{ism} in all regions observed in this merger has smoother structure than disc galaxies, at the same physical sizes.

\subsection{Gas dynamics}
Figure~\ref{fig:1d_alpha_vir} shows a significant difference in median virial parameters measured in \ngc{} and the \gls{phangs} sample. Lines of constant virial parameter in Figure~\ref{fig:2d_KDE_vd_vs_sd} show how those differing virial parameters result in an offset between the two samples, potentially originating from a real difference in the dynamical states of the molecular gas.

The majority of gas across the \gls{phangs} galaxies is nearly self-gravitating or slightly over-pressurized across almost \SI{2}{\dex} in surface density and \SI{1}{\dex} in velocity dispersion. The gas in \ngc{} instead has enough excess velocity dispersion over about \SI{\sim 1}{\dex} in both surface density and dispersion that the gas appears to be nowhere near self-gravitating. However, both of these interpretations of the virial parameters do not include the contribution of external pressure confinement in estimating the boundedness of the gas in clouds.

\cite{Sun2020} conclude that the contribution of external weight from atomic, stellar, and molecular mass in a clumpy \gls{ism} can balance the over-pressurization seen in the \gls{phangs} sample and bring much of the gas into being bound in \glspl{gmc}. Assuming the high virial parameters estimated in \ngc{} are solely from the omission of external pressure, we can estimate the external mass surface densities required for pressure equilibrium within each beam. We find there would have to be between \SIlist{65;4200}{\solarmass\per\square\parsec} in external mass confining the molecular clouds in the non-nuclear pixels of \ngc{}, and up to \SI{1e4}{\solarmass\per\square\parsec} for some pixels in the nuclei. We summarize our method in Appendix~\ref{external_material_appendix}.

Stellar mass surface densities of nearby spirals can cover much of this mass range from \SIrange{\sim 60}{2000}{\solarmass\per\square\parsec} \citep{Ler2008,Jim2019}. \cite{Eng2003} measured \SI{\sim 1e9}{\solarmass} of atomic gas in the central \SI{5}{\kilo\parsec} of \ngc{}, giving an average atomic gas surface density of about \SI{51}{\solarmass\per\square\parsec}. Including an estimate here for the contribution from large-scale and diffuse molecular gas external to the densest regions on \gls{gmc} scales is not straightforward with an analysis like that of \cite{Sun2020}. However, from these estimates it seems plausible that a portion of the densest molecular gas we observe in \ngc{} is being pressure confined by external matter. However, the highest dispersion gas likely does not have enough external pressure to be in pressure equilibrium with the surrounding material.

Estimates of pressures surrounding \glspl{gmc} in systems like \ngc{} have been made previously. HI cloud collisions have been estimated to produce localized pressures of \SI{\sim 1e8}{\kelvin\per\cubic\centi\metre} but result in average ambient pressures of only \SI{\sim 5.5e5}{\kelvin\per\cubic\centi\metre} \citep{Jog1992}. \cite{Wil2003} estimate external pressures for supergiant molecular complexes in NGC~4038/9 due to the hot \gls{ism} to be on average \SI{6e5}{\kelvin\per\cubic\centi\metre}. They further argue that the low X-ray filling factor results in their pressure estimate being consistent with the localized pressure of \cite{Jog1992}. Compared to \ngc{}, the localized pressures from \cite{Jog1992} exceed the non-nuclear pressures by an order of magnitude and are consistent with the pressures measured within the central \SI{1}{\kilo\parsec} of the nuclei. However, the ambient pressures in these models are too low by two to three orders of magnitude. \cite{Jog1992} explain that \glspl{gmc} may be subject to the localized pressure if they are uniformly surrounded by and in direct contact with several HI clouds. Given the difference of up to an order of magnitude between the HI and molecular gas surface densities in \ngc{}, it is unlikely the atomic component is the dominant source of pressure confinement.

\cite{Joh2015} estimated the external pressure on the Firecracker molecular cloud in NGC~4038/9 from the overlying molecular material to be \SI{\sim 1e7}{\kelvin\per\cubic\centi\metre}. Comparing this to the requisite external pressure of \SI{\sim 1e9}{\kelvin\per\cubic\centi\metre} needed to explain the velocity dispersion of the gas showed additional sources of pressure (like cloud-cloud collisions) were needed. Given the similar internal pressures but lower surface densities in \ngc{} compared to NGC~4038/9, it is even less likely that external molecular gas could gravitationally bind the gas in \ngc{}. \cite{Tsu2019,Tsu2020} estimated external pressures of \SIrange{\sim 1e8}{1e9}{\kelvin\per\cubic\centi\metre} due to molecular cloud collisions for the five giant molecular complexes in NGC~4038/9. Similar external pressures could account for some of the highest internal pressures in \ngc{}, but would beg the question of whether there is enough molecular gas in the central \SI{6}{\kilo\parsec} of \ngc{} undergoing cloud-cloud collisions to produce the pressures estimated here.

Returning to the offset of \ngc{} relative to the \gls{phangs} sample in Figure~\ref{fig:2d_KDE_vd_vs_sd}, \ngc{} makes it clear that the velocity dispersion-surface density parameter space does not capture the full picture of the dynamics of molecular gas. The tight correlation seen across the \gls{phangs} main sample may be an artifact of selecting dynamically similar galaxies that are massive, disc-dominated, and star forming.

Related to this is the actual source of the turbulence. \cite{Kru2018} argue that the only way to power dispersions $\gtrsim$\SI{10}{\kilo\metre\per\second} is through bulk gas flows that are ubiquitous in interacting galaxies. Differences in the power sources for turbulence between interacting and disc galaxies may be manifesting themselves in the state of the gas as probed by Figure~\ref{fig:2d_KDE_vd_vs_sd}. It is not trivial to trace the sources of turbulence back from the gas dynamics in this way however since NGC~4038/9 lies along the \gls{phangs} main sample relation while \ngc{} is offset from it. The individual properties of each interacting system may also change this offset, and the offset could change with time as the interactions evolve.

\subsection{Virial diameter}
\begin{figure}
\includegraphics{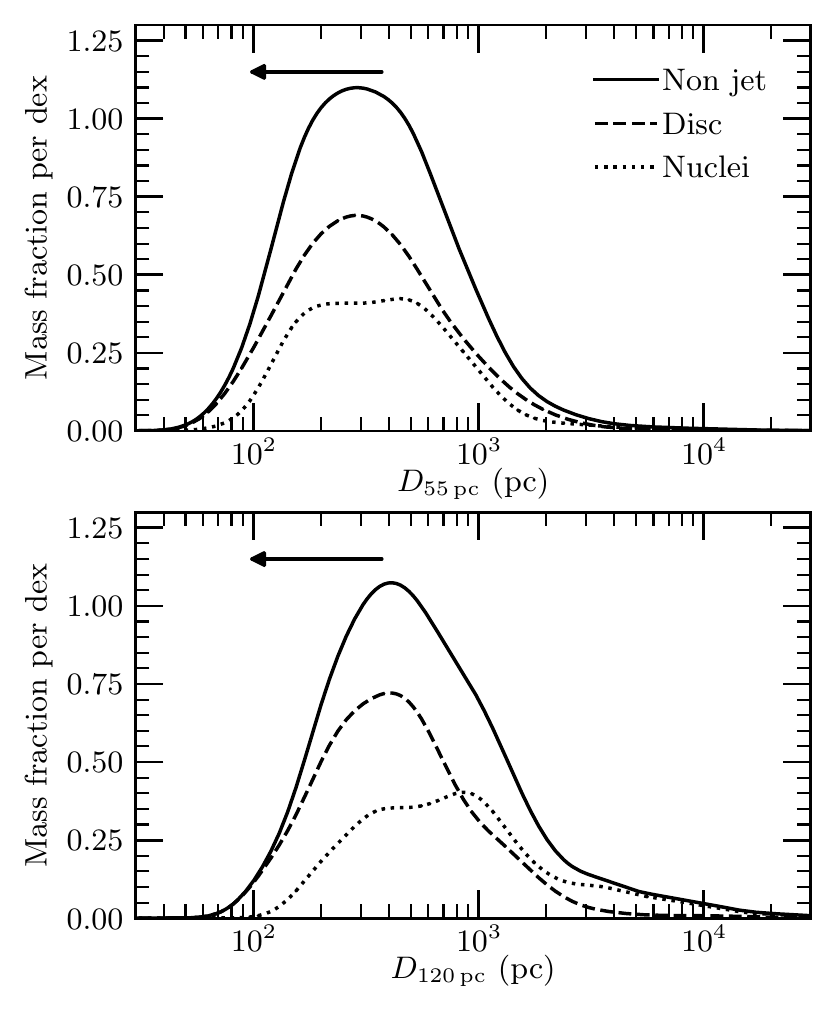}
\caption{Cloud diameters calculated assuming a virial parameter of \num{2} for each pixel where surface density and velocity dispersion were measured in \ngc{}. Linestyles are the same as in Figure~\ref{fig:1d_sd}. Arrows show how values would shift if we adopt the Milky Way \gls{co} conversion factor of \SI{6.25}{\solarmass\per\square\parsec}(\si{\kelvin\kilo\metre\per\second})$^{-1}$ instead of the \gls{lirg} conversion factor of \SI{1.38}{\solarmass\per\square\parsec}(\si{\kelvin\kilo\metre\per\second})$^{-1}$. }
\label{fig:1d_KDE_virial_diameter}
\end{figure}

Instead of asking if the gas is bound or not at the observed scales, we can ask at what size would the gas be in approximate free-fall collapse ($\alpha_{\mathrm{vir}}=2$), given the measured surface densities and velocity dispersions. Figure~\ref{fig:1d_KDE_virial_diameter} shows the Gaussian \glspl{kde} for these sizes. The all-pixel distributions peak around \SIrange{300}{400}{\parsec} between \num{55} and \SI{120}{\parsec} resolution, respectively. Given the diameter at the peak of the distributions is roughly five beams across at \SI{55}{\parsec} resolution, a cloud-decomposition analysis would be able to identify if most of the gas is grouped into clouds of this size.

If this interpretation is correct then it would imply that the molecular gas is mostly gravitationally bound, but on quite large scales of a few hundred parsecs. One alternative could be that the velocity dispersions are biased towards high values by multiple clouds along the line of sight, and removing that effect would drive these sizes down. It is also possible that even at our highest resolution there are multiple clouds within the beam and that the cloud-to-cloud velocity differences are driving up the dispersion (and the sizes in Figure~\ref{fig:1d_KDE_virial_diameter}). Without a good understanding of the cloud size and mass distributions in mergers like \ngc{}, this later alternative can only be explored with higher spatial resolution observations. Finally, it could be that the molecular gas is simply not bound at these scales and is behaving more similarly to the atomic gas traced by HI in spiral galaxies.

\subsection{Expectations for a merger}
High molecular gas surface densities, linked to enhanced quantities of molecular gas in interacting and merging galaxies, have been observed in many systems previously \citep[e.g.][]{Bra1993,Kan2013,Vio2018}. This has also been reproduced in numerical studies. \cite{Mor2019}, for example, measure a \num{\sim 18} per cent increase in cold-dense gas ($n = \SIrange[range-phrase=-]{10}{1000}{\per\cubic\centi\metre}$) and a \num{\sim 240} per cent increase in cold ultra-dense gas ($n > \SI{1000}{\per\cubic\centi\metre}$).

A possible origin of the extreme velocity dispersions in \ngc{} is that the star formation rate was enhanced first, which then increased the velocity dispersion through stellar and supernova feedback. However, theoretical and numerical investigations into the sources of turbulent energy indicate that even at mass surface densities and turbulence levels replicating \gls{u/lirg} conditions and at the maximal limit of star formation rates for discs, the gas velocity dispersion can only reach \SI{\sim 10}{\kilo\metre\per\second} \citep{She2012,Kru2018}. Gas inflow was proposed by \cite{Kru2018} as the main mechanism to drive velocity dispersions above that limit. So while a portion of the measured velocity dispersion distribution can originate from star formation feedback, it is possible the highest dispersions are driven by gas flows.

Since the molecular gas at the scales observed in \ngc{} appears smooth in comparison to disc galaxies and extreme external pressures are required to bind the gas, what are the actual sites of star formation like in this merger? It would seem that self-gravitating and collapsing overdensities must exist at scales smaller than \SI{55}{\parsec}, potentially like Galactic clumps and cores. \cite{Mor2019} report that interactions dramatically increase the mass fraction of the densest cold gas in their simulations, indicating the source of enhanced star formation rates. However, this densest component only makes up \num{\sim 0.2} percent of the cold gas mass. Overall they measure an enhancement in diffuse molecular gas that cannot collapse. These models could imply that we are observing the high surface densities of the enhanced cold gas reservoir, but only at the scales where the gas is too diffuse to directly contribute to star formation. Marginally-bound structures at the highest densities are driven to collapse by the interaction and thus the star formation rate increases, but the addition of turbulent energy from inflowing gas will act to over-pressurize the bulk of the molecular gas.

\subsection{Caveats}\label{caveats}
We first consider the applicability of using a \gls{co} conversion factor to estimate molecular gas masses from integrated \gls{co} line intensities. Estimating mass from the optically thick \gls{co} emission with a conversion factor relies on the presence of virialized clouds which can relate their emitting surface area to the total mass in their volume. While the molecular gas in \ngc{} is estimated to be well out of virial equilibrium due to self-gravity alone, it may be virialized when external pressure and gravitational terms (e.g. stars and the overall galactic potential) are included. If the molecular gas appears smooth on \SI{55}{\parsec} scales because the majority of clouds are smaller than the beam, then the integrated line intensities would be proportional to the number of clouds within the beam. In lieu of optically thin emission, the collection of clouds within each beam may act optically thin such that each cloud contributes an average mass so the integrated intensity is still proportional to the total mass within the beam \citep[the so-called ``mist'' model,][]{Bol2013}. Finally, beyond considering individual virialized clouds, the \gls{u/lirg} conversion factor used here has also been derived in the centres of \glspl{u/lirg} using \gls{co} spectra to estimate dynamical masses assuming smoothly distributed molecular gas \citep{Dow1998}.

The main assumption to be concerned with in the mist model when applying it to a system like \ngc{} is that the sub-beam structure cannot be overlapping in spatial-spectral space. The high surface densities mean there is likely to be some degree of gas structures overlapping along the line of sight spatially, but the gas would also have to be Doppler shifted to the same velocities for the molecular mass to be underestimated. For this effect to entirely account for the offset in Figure~\ref{fig:2d_KDE_vd_vs_sd} would require \num{\sim 4.5} clouds both along the line of sight and at the same velocity in most pixels. An average of \num{4.5} clouds along each line of sight seems excessively large.

We now consider the value we chose for the conversion factor. Molecular gas properties potentially present in \ngc{} such as higher temperatures, larger velocity dispersions, low \gls{co} optical depths, and non-virialized gas have all been argued as sources of low conversion factors in various studies of \glspl{u/lirg} \cite[e.g.][]{Wil1992, Pap1999, Zhu2003}. However, by switching to the Galactic value, the surface densities would increase by a factor of \num{\sim 4.5}. This would result in the \ngc{} non-nuclear sample overlapping considerably with the sample from NGC~4038/9, and it would also shift the bulk of the virial parameter distribution to \num{\sim 2}. Gas at these densities and scales being bound by self-gravity would call into question the findings of \cite{Mor2019} where they argue the majority of the enhanced molecular gas is not participating in star formation. It could instead be that the offset of \ngc{} is real and that NGC~4038/9 would be better described by the \gls{u/lirg} conversion factor which would shift NGC~4038/9 closer to where \ngc{} appears in Figure~\ref{fig:2d_KDE_vd_vs_sd}. However, measurements of the conversion factor in NGC~4038/9 have actually come in either consistent with the Galactic value or even higher \citep{Wil2003,Sch2014}.

Differences in the proper conversion factor between \ngc{} and NGC~4038/9 could originate from their differing merger stages as well as the regions analyzed. Our observations of \ngc{} cover the very centre of the system, where the conversion factor may be shifted to lower values \citep[e.g.][]{Wei2001,Hit2008,Sli2013}. It is also likely that the conversion factor varies significantly across different regions in this merger as \cite{Ren2019} measure variations up to a factor of \num{2.2} in their simulation of a NGC~4038/9-like merger. They also expect spatial variations are even greater since they focused their analysis on actively star-forming regions. Directly estimating the conversion factor may be possible with these data but requires cloud decomposition in spatial and spectral dimensions. This will be investigated in a future paper.

Caution should be used when interpreting the velocity dispersions measured in both \ngc{} and NGC~4038/9 due to the complex morphologies of interacting galaxies which makes it much more likely to observe multiple cloud structures along the line of sight. This is a limitation of the current pixel-based analysis. The degree to which both mergers, and even each region within the two systems, are affected by cloud multiplicity is likely to result in a non-trivial rearrangement of their distributions in Figure~\ref{fig:2d_KDE_vd_vs_sd}. Our future cloud-decomposition analysis of \ngc{} will help to eliminate this effect. We also note that while we attribute the spectral line widths to turbulence in our discussions, unresolved bulk motions also contribute to the measured dispersions. Given the high spatial resolution our observations however, an estimate of galactic rotation from the extrema of moment \num{1} maps contributes at most \SI{5.5}{\kilo\metre\per\second} of dispersion across a single \SI{120}{\parsec} beam. Bulk flow of gas from the merging process is also likely contributing in some regions of \ngc{}.

Interpretation of the virial parameters and internal turbulent pressures calculated for the mergers should also be made with caution. If the underlying assumptions are satisfied in \ngc{} as well as in all of the galaxies presented by \cite{Sun2018}, then using the same expressions for calculating these quantities should make the comparison of results straightforward. For the virial parameter, these assumptions include spherical clouds with density profiles of $\rho \propto r^{-2}$ that are the same size as the beam. The internal pressure relies on each beam being filled by one virialized cloud along each line of sight \citep{Ber1992}. Based on many previous studies of \gls{gmc} properties in disc galaxies, these conditions are likely to hold in the main \gls{phangs} sample. However, without even higher resolution observations we cannot determine if these conditions hold in \ngc{}. The virial parameter distribution well above one indicates that the internal pressure would be underestimated, since the gas kinetic energy associated with bulk motions (like expansion) would not be accounted for in calculating the pressure. However, lines of sight with multiple components would overestimate the dispersion in the average cloud leading to an overestimation of the internal pressures.

\section{Conclusions}
\label{conclusions}
We have observed the central \SI{6}{\kilo\parsec} of the closest \gls{lirg}, \ngc{}, in \gls{co} (\num{2}--\num{1}) with \gls{alma} at \gls{gmc}-scale resolution to obtain pixel-by-pixel distributions of mass surface density, velocity dispersion, peak brightness temperature, virial parameter, and internal turbulent pressure. From these distributions we find:
\begin{itemize}
\item Assuming the \gls{u/lirg} conversion factor of \SI{1.38}{\solarmass\per\square\parsec}(\si{\kelvin\kilo\metre\per\second})$^{-1}$, molecular mass surface densities range between \SIrange{8}{5500}{\solarmass\per\square\parsec}. This range is similar to the overlap region of NGC~4038/9, but extends to lower surface densities. \ngc{} has surface densities above the majority of the disc measurements from \gls{phangs} reported by \cite{Sun2018}.

\item Velocity dispersions range from \SIrange{10}{200}{\kilo\metre\per\second} in \ngc{}. The median outside the central kiloparsec radius is \SI{25}{\kilo\metre\per\second}. These velocity dispersions are well above the majority of the \gls{phangs} disc galaxies, and are consistent with NGC~4038/9.

\item The vast majority of the gas appears unbound at \SI{55}{\parsec} scales with virial parameters above \num{3}, and medians even reaching \numrange{7}{19}. Coupled to this, we estimate very high turbulent pressures from \SIrange{1e5}{1e10}{\kelvin\per\cubic\centi\metre}. External pressure on \glspl{gmc} may bind a fraction of the gas, but there is still likely a significant fraction of mass that is unbound.

\item \num{50} per cent of the mass in the non-nuclear pixels of \ngc{} has peak brightness temperatures significantly above almost all measurements in the main \gls{phangs} sample as well as the overlap region of NGC~4038/9. All measurements in the nuclei of \ngc{} are above those in all other galaxies analyzed here. This suggests the majority of molecular gas in \ngc{} is significantly warmer than disc galaxies and even the most vigorously star forming region of NGC~4038/9.

\item The observation of little or no trend in surface density, velocity dispersion, and peak brightness temperature with resolution indicates the molecular medium has a smoother structure at \SI{55}{\parsec} scales than in the \gls{phangs} disc galaxies. A smooth molecular \gls{ism} appearing unbound on these scales could be consistent with FIRE-2 numerical results \citep{Mor2019}. It may also be that gas at these scales in \ngc{} more closely resembles the dynamics of giant molecular associations in disc galaxies.
\end{itemize}

An additional test of the boundedness of the gas in \ngc{} will be carried out using a spatial and spectral decomposition of the \gls{co} emission in a future paper. Instead of assuming the beam size represents the relevant scale,this analysis will produce empirical estimates of the sizes of the molecular structures. We will also attempt to estimate the \gls{co} conversion factor from the cloud size and linewidth catalog. However, the virial parameter estimates reported in this work make it likely that this approach will produce at best an upper limit. Independent methods to estimate the conversion factor (e.g. estimating dust masses from the continuum or optically thin molecular line isotopologues) should also be investigated. 

Applying both pixel-by-pixel and cloud-decomposition analyses to more merger systems at this resolution will help to determine if the apparent \gls{ism} dynamics and structure in \ngc{} are general properties of these extreme systems. It can also further test the universality of gas dynamical state reported by \cite{Sun2018} across the \gls{phangs} main sample and overlap region in NGC~4038/9. In a forthcoming paper we will apply these analyses, at the same physical scales, to the entire NGC~4038/9 merger system in \gls{co} (\num{2}--\num{1}). This will also allow us to probe the nuclear regions for similarities with the nuclei of other galaxies.

Finally, higher resolution imaging of \ngc{} in \gls{co} (\num{2}--\num{1}) would explore at what scale the molecular gas goes from a smooth medium to the clumpy one we assume must exist due to the presence of significant star formation. \gls{alma} can observe down to \SI{\approx 9}{\parsec} scales to search for significant changes in the physical quantities discussed here. Any trends with resolution would indicate a structured \gls{ism} like what is seen in the \gls{phangs} galaxies at \SI{45}{\parsec}. Whether we will find clouds on scales smaller than \SI{55}{\parsec} in \ngc{} that are analogous to \glspl{gmc} in nearby discs, or that the molecular gas remains unbound down to scales and densities that are more similar to Galactic clumps or cores is a question that awaits maximal resolution data from \gls{alma}.

\section*{Data availability}
This paper makes use of the following ALMA data: ADS/JAO.ALMA\#2015.1.00714.S (accessed from the \gls{alma} Science portal at \url{almascience.org}). ALMA is a partnership of ESO (representing its member states), NSF (USA) and NINS (Japan), together with NRC (Canada), MOST and ASIAA (Taiwan), and KASI (Republic of Korea), in cooperation with the Republic of Chile. The Joint ALMA Observatory is operated by ESO, AUI/NRAO and NAOJ. The National Radio Astronomy Observatory is a facility of the National Science Foundation operated under cooperative agreement by Associated Universities, Inc.

The derived data generated in this research will be shared on reasonable request to the corresponding author.

\section*{Acknowledgements}

N.B. wishes to thank the \gls{naasc} for hosting him on a data reduction visit, and Amanda Kepley and Adam Leroy for helpful discussions on imaging this complex data set. We thank Jiayi Sun for access to and discussion of his thresholding code. We thank Kazushi Sakamoto for helpful comments on this paper. We thank the anonymous referee for detailed comments that improved the content of this paper. 


The research of C.D.W. is supported by grants from the Natural Sciences and Engineering Research Council of Canada and the Canada Research Chairs program.

This research made use of \textsc{Astropy}, a community-developed core \textsc{Python} package for Astronomy \citep[\url{http://www.astropy.org},][]{astropy2013,astropy2018}. This research also made use of the \textsc{Matplotlib} \citep{Hun2007}, \textsc{NumPy} \citep{van2002}, \textsc{scikit-learn} \citep{Ped2011}, \textsc{Jupyter Notebook} \citep{Klu2016}, and \textsc{statsmodels} \citep{Sea2010} \textsc{Python} packages. This research has made use of NASA’s Astrophysics Data System. This research has made use of the VizieR catalogue access tool \citep{Och2000}. This research has made use of the NASA/IPAC Extragalactic Database (NED), which is funded by the National Aeronautics and Space Administration and operated by the California Institute of Technology. This research has made use of the SIMBAD database, operated at CDS, Strasbourg, France \citep{Wen2000}.




\bibliographystyle{mnras}
\bibliography{references}



\appendix
\renewcommand{\thesubsection}{\Alph{section}}
\setcounter{section}{0}

\section{Estimating required external material}
\label{external_material_appendix}
The virial parameters presented in Section~\ref{vir_results} neglect the effect of external pressure on the cloud. We make a rough estimate of the pressure required to balance the internal turbulent pressure following a simplified form of that by \cite{Sun2020}. We aim to satisfy the criterion that the external pressure, $P_{\mathrm{ext}}$, is equal to the internal pressure, $P_{\mathrm{turb}}$. A simple expression for the gravitational pressure from mass external to the cloud is
\begin{equation}
P_{\mathrm{ext}} k_{B} = CG\Sigma_{\mathrm{cloud}}\Sigma_{\mathrm{tot}}
\end{equation}
where $k_{B}$ is the Boltzmann constant, $C$ is a constant that contains normalizations for the assumed geometries of all of the material, $G$ is the gravitational constant, $\Sigma_{\mathrm{cloud}}$ is the mass surface density of the molecular cloud, and $\Sigma_{\mathrm{tot}}$ is the surface density of all material along the line of sight. 

Neglecting dark matter in the inner portion of this merger, $\Sigma_{\mathrm{tot}}$ is made up of
\begin{align}
\Sigma_{\mathrm{tot}} &= \Sigma_{\mathrm{cloud}} + \Sigma_{\mathrm{ext,mol}} + \Sigma_{\star} + \Sigma_{\mathrm{atom}} \\
                                    &= \Sigma_{\mathrm{cloud}} + \Sigma_{\mathrm{ext}}
\end{align}
where $\Sigma_{\mathrm{ext,mol}}$ is the molecular gas surface density outside the cloud, $\Sigma_{\star}$ is the stellar surface density, $\Sigma_{\mathrm{atom}}$ is the atomic surface density, and $\Sigma_{\mathrm{ext}}$ is the sum of the components external to the cloud. Rewriting the external pressure with these separate terms gives
\begin{equation}
P_{\mathrm{ext}} k_{B} = G\Sigma_{\mathrm{cloud}}\left(C_{1}\Sigma_{\mathrm{cloud}} + C_{2}\Sigma_{\mathrm{ext}}\right) \label{eq:external_pressure}
\end{equation}
where $C_{1}\Sigma_{\mathrm{cloud}}$ accounts for the self-gravity of the cloud and $C_{2}\Sigma_{\mathrm{ext}}$ the weight of material external to the cloud. $C_{1}$ and $C_{2}$ are again constants that depend on the geometry of these components. Assuming constant density within a spherical cloud, $C_{1} = 3\pi /8$ \citep[e.g. Equation~A2 from][]{Sun2020}.

Setting the external pressure equal to the measured internal pressure and rearranging Equation~\ref{eq:external_pressure} we have an expression for the external mass surface density required to balance the internal turbulent pressure
\begin{equation}
\Sigma_{\mathrm{ext}} = \frac{1}{C_{2}}\left(\frac{k_{B} P_{\mathrm{turb}}}{G \Sigma_{\mathrm{cloud}}} - \frac{3\pi}{8}\Sigma_{\mathrm{cloud}}\right).
\end{equation}
If we treat $C_{2}$ as tunable between $3\pi /8$ (spherical) and $\pi /2$ (disc with scale height much greater than the molecular disc scale height), then we can explore a reasonable range of external pressures given different combinations of relative geometries of the multiple components. This was the most useful approach considering that $C_{2}$ must encode the volumetric average of all three external components relative to the molecular clouds in a morphologically disturbed merger.

Calculating external surface densities, pixel-by-pixel using our measurements at \SI{55}{\parsec} resolution, gives disc mass-weighted \numlist{16;50;84} percentiles of $(\text{\SIlist{100;1000;3000}{\solarmass\per\square\parsec}}) / C_{2}$. The combined-nuclei percentiles are $(\text{\SIlist{2e3;6e3;20e3}{\solarmass\per\square\parsec}}) / C_{2}$. The same calculation for the \gls{phangs} sample results in most pixels having external surface densities of zero or less, which we interpret as the majority of gas being self-gravity dominated \citep{Sun2020}. The remaining pixels in the \gls{phangs} sample appear around $(\text{\SIrange{50}{100}{\solarmass\per\square\parsec}}) / C_{2}$.

An important component this model neglects is the pressure contribution from self-gravity of the external material \citep[e.g. the atomic term in Equation~A9 from ][]{Sun2020}. This primarily depends on the surface density of the atomic gas, and acts to reduce the external surface density required to confine the molecular cloud. Galaxies in the \gls{phangs} sample are strongly impacted by this term, and so the full treatment by \cite{Sun2020} is recommended to assess their pressure equilibrium. We do not expect the atomic surface density to provide a very substantial correction in a merger. Since we estimate \ngc{} requires significantly higher external surface densities than the \gls{phangs} galaxies, it is unlikely the missing external self-gravity term would drastically alter our results.


\bsp	
\label{lastpage}
\end{document}